\documentclass[preprint,12pt,authoryear]{elsarticle}

\usepackage[utf8]{inputenc}
\usepackage[T1]{fontenc}
\usepackage{lmodern}
\usepackage{amsmath}
\usepackage{booktabs}
\usepackage{array}
\usepackage{adjustbox}
\usepackage{graphicx}
\usepackage{pgfplots}
\pgfplotsset{compat=1.18}
\usetikzlibrary{arrows.meta}

\definecolor{atcdark}{HTML}{184F95}
\definecolor{atclight}{HTML}{6DA7EC}
\definecolor{atctint}{HTML}{E3EEFC}

\DeclareUnicodeCharacter{2212}{\ensuremath{-}}

\makeatletter
\def\ps@pprintTitle{%
  \let\@oddhead\@empty
  \let\@evenhead\@empty
  \def\@oddfoot{\footnotesize\itshape Preprint\hfill}%
  \let\@evenfoot\@oddfoot}
\makeatother

\begin{document}

\begin{frontmatter}

\title{Explain Your Code: AI-Based Examination at Class Scale}

\author[a]{David Sarne}
\author[a]{Shachar Sagi}
\author[b]{Yossi Ben Zion\corref{cor1}}
\ead{Yossi.Ben-zion@biu.ac.il}
\cortext[cor1]{Corresponding author.}

\affiliation[a]{organization={Department of Computer Science, Bar-Ilan University},
  city={Ramat Gan},
  postcode={5290002},
  country={Israel}}
\affiliation[b]{organization={Department of Physics, Bar-Ilan University},
  city={Ramat Gan},
  postcode={5290002},
  country={Israel}}

\begin{abstract}
Generative AI has weakened the link between working code and a student's understanding of it. Oral examination, the traditional test of that understanding, does not scale to large courses. We report the full-course deployment of a three-agent AI system that examines students on their own code. The first, the examiner, asks three questions anchored to specific lines of the submission, adapting difficulty to each answer, in a 16-minute text dialogue. The other two grade: a code reviewer that never sees the examination and a grader without direct access to the code. Authorship is assessed separately and is advisory. The system examined 247 students in an operating systems course (three exercises, 546 sittings, two examination languages). The grader's oral score was uncorrelated with the same submission's code quality score (Spearman $\rho$ = −0.073, 95\% CI [−0.205, 0.063], n = 216) yet tracked degree grade point average ($\rho$ = 0.419, [0.294, 0.523], n = 214), while automated test scores sat at the ceiling. The mean of a student's two oral scores correlated with grade point average at 0.468 ([0.353, 0.568], n = 210), within the range of grades in three single-semester courses built on programming assignments. Model spend averaged at most \$1.37 per sitting. In surveys, students began skeptical of the AI examination and rated it more favorably after sitting it twice. These results indicate that examining students on their own code is feasible at class scale, and that its score relates to academic standing where the code quality score does not.
\end{abstract}


\end{frontmatter}

\section{Introduction}\label{sec:introduction}

By students' own accounts, generative AI is now a routine part of producing assessed work: in successive surveys of UK undergraduates, 88\% in 2025 and 94\% in 2026 reported using it to help with their assessments \citep{freeman2025,stephenson2026}. In programming, where a language model can produce a working solution from a prompt in seconds, a passing submission stops being clear evidence that the student who handed it in can do the work. Computer science undergraduates have been observed describing a task to a model, checking that the returned code runs and submitting it, without working through the errors and revisions that build understanding \citep{hak2025}. In a randomized experiment with developers learning an unfamiliar Python library, AI assistance impaired conceptual understanding, code reading and debugging without significant efficiency gains on average \citep{shen2026}. The cost does not fall on all students equally. Novices who did not struggle used generative AI to accelerate work they had already intended to write, while those who struggled finished with an illusion of competence \citep{prather2024}. The change is visible in what students submit as well. On a single assignment left unchanged for five years, students after the release of ChatGPT changed more code between consecutive submissions but gained less score from each, which the authors read as a loss of both productivity and learning \citep{zhang2026}. The gap predates generative AI: about a third of introductory students whose programs passed the automated tests struggled to explain their own code \citep{lehtinen2021struggle}. Assessment that reads only the artifact therefore rests on an inference, from working code to understanding, that these findings put in doubt.

One response is the oral examination, in which a student defends their work live with no tools at hand. It is both used and advocated as an integrity response: institutional guidance routes suspect submissions to it, and researchers argue that written assessment may need replacing with face-to-face forms \citep{kofinas2025}. A human oral defense for every student in a large course, however, is expensive, and hard to keep consistent across graders. An automated system deployed in an undergraduate artificial intelligence and machine learning course at NYU Stern, across cohorts of 36 and 37 students, shows that AI can conduct oral assessment at very low cost \citep{ipeirotis2026}. Its authors also report that the model panel graded more strictly than human evaluators and agreed poorly until a consultation step. That system questions the student from an instructor rubric and a text description of the project rather than from the submitted artifact. Section~\ref{sec:related} reviews three lines of prior work: systems that ask questions anchored to a student's own code but do not examine, systems that examine autonomously and adaptively but do not read the submission, and designs that combine the two but have been described without participants or deployment \citep{church2025viva,vintila2024avert,frankford2026chatbot}. To our knowledge, questioning from the submitted code and autonomous examination have not been brought together in a running course.

\begin{figure}[t]
\centering
\includegraphics[width=\linewidth]{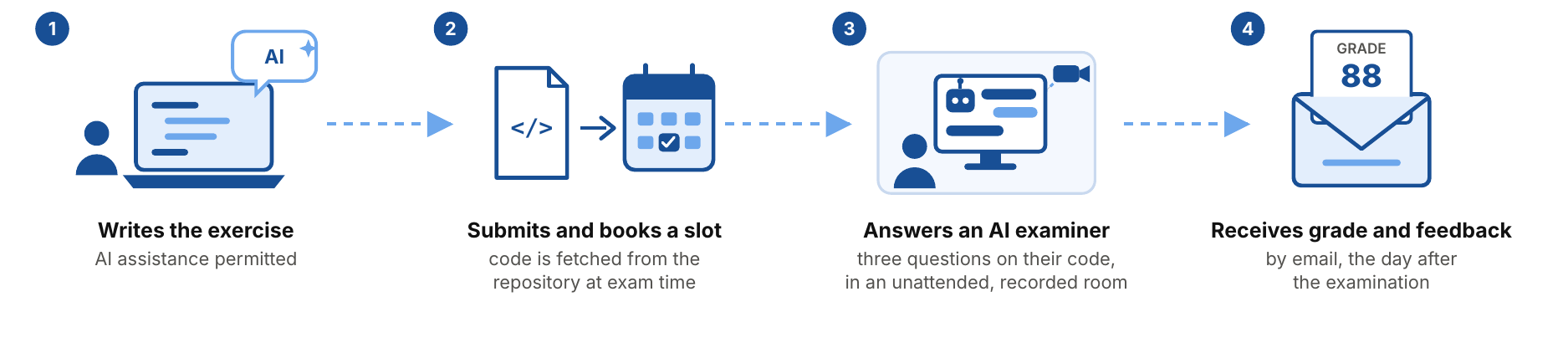}
\caption{The examination from the student's side. The student writes the code for the exercise, with AI assistance permitted, submits it through a repository and books an examination slot. In an unattended, recorded room, the student answers in writing three questions from the AI examiner about their own code. The grade and written feedback arrive by email on the following day. The grade shown is illustrative.}
\label{fig:journey}
\end{figure}

We report on ATC, the Autonomous Testing Center, an automated examination system for programming assignments that examines each student autonomously on their own submitted code. The system is built from three agents, each a large language model given its own instructions and its own inputs. The first, the examiner, reads the student's own submitted code and questions the student on it. Its questions are anchored to specific lines of the code and adapt in difficulty to each answer, and it scores each answer as it is given, which guides the difficulty of the next question. The second, the Code Reviewer, sees the code but not the examination, and produces a code quality score. The third, the Grader, scores the examination as a whole from its transcript, without direct access to the code, and produces an oral score. The final grade combines the oral score and the code quality score by fixed arithmetic. Understanding and authorship, meaning whether the student appears to have written the code, are kept as separate axes, so that a nervous but genuine student is not treated as an impostor. The examination is conducted in text: the student reads the questions on screen and answers in writing. We keep the term oral, as is usual for an examination in which a student defends their own work in dialogue, although every answer here was typed. We chose text because a typed answer lets a student organize an explanation of code and revise it before submitting it, and it avoids transcribing identifiers and technical vocabulary from speech. We deployed the system on the Introduction to Operating Systems course at Bar-Ilan University, in two examination languages (Hebrew and English). Figure~\ref{fig:journey} shows the process from the student's side.

Our aim is feasibility rather than efficacy: we ask whether such a system can run on a full course at the cost we report in Section~\ref{sec:cost} and produce grades that relate to other measures as the design predicts. The course also graded programs with automatic test suites, which run the submitted program against fixed test cases and produce an automated test score. Before any examination data were extracted for analysis, we fixed five success criteria, each with a numeric threshold: three on the examination scores and two on the automated test scores. We report the outcome of each, and a criterion that failed is reported as a finding rather than revised after the fact. Analyses that were not pre-specified are reported as exploratory. We do not claim that these grades are validated against human examiners. That comparison is left to future work.

This paper makes four contributions. First, submission-aware questioning inside an examination. The questions are generated from the student's own code and anchored to specific lines, rather than to a rubric or a project summary. They are asked adaptively in dialogue, and the examination produces a grade of record, one that counts toward the course grade. Second, a three-agent assessment design in which each agent sees only part of the evidence. The code review is kept apart from the examination, their scores are combined arithmetically, and understanding and authorship are scored independently. The examiner also carries fairness and safety behaviors: a distress path that ends the examination when a student signals distress, handling of instructions planted in an answer to manipulate the examiner (prompt injection), explicit rules against penalizing trivial slips, and a complaint flow for a student shown code that is not theirs. The distress path, the injection flag and the complaint flow were in place throughout, but were not triggered during the deployment. Third, a deployment at full class scale, across three exercises in two examination languages, at an average model cost of at most \$1.37 per sitting (Section~\ref{sec:cost}). Fourth, empirical observations against criteria fixed in advance. The oral score is uncorrelated with the code quality score of the same submission while tracking degree grade point average. The automated test scores sit at a ceiling, which leaves no room to detect whether students who expected automatic grading wrote their code differently. An exploratory comparison places two short oral examinations, as an indicator of academic standing, within the range of grades in single-semester courses that, like this one, are built on programming assignments.

\section{Related work}\label{sec:related}

The previous section argued that a working submission is weak evidence of student understanding, and that oral examination is the response the integrity literature reaches for. In an oral defense, the examiner questions the student on the work they handed in and follows up on each answer. This section therefore asks which of the systems built to put that response into software have two properties, one for what is asked and one for how. The first is that the questions come from the student's own submission, rather than from a rubric, a project description or the course materials, so that they test the student's account of it. The second is that the system itself conducts an adaptive dialogue about that work, rather than issuing fixed items or handing questions to a human examiner. Only a dialogue can pursue a weak answer, and only one conducted without an examiner can run at scale. How far each system has been deployed is stated in Table~\ref{tab:related} but is not used to sort the systems.

Assessment of programming assignments has mostly rested on a different instrument: automatic execution of the submitted program against test cases, which has been surveyed as a field in its own right \citep{douce2005testbased} and as part of automated assessment more broadly \citep{paiva2022review}. The grade a test-based system produces varies with the test suite used \citep{clegg2020testsuite}. Because such tests check only what a program outputs, assessment that also examines the code itself has been proposed \citep{jin2020blackbox}, and the strengths and limitations of existing systems have been reviewed \citep{ullah2018effect}. Section~\ref{sec:observed} reports a ceiling in this instrument's scores in the present deployment.

Oral examination in computing is also more than an integrity response. It has been studied as an assessment method in its own right: in introductory computer science \citep{ohmann2019oral}, in software courses at several institutions \citep{ohmann2025multiinstitutional}, and as interviews on individual take-home programming assignments \citep{kannam2025code}. In an introductory course of 628 students, 21 staff gave each student two five-minute oral examinations, at a cost of about 180 staff hours \citep{ellis2026scale}. \citet{motallebi2026map} map this work systematically for computing. In all of these studies people conduct the examination, which is why they are not in Table~\ref{tab:related}.

\begin{table}[!tbp]
\centering
\small
\caption{Systems that question students about their work, classified by the two properties used in this section. Deployment figures are as reported in each paper. For ATC, a sitting is one student's attempt at one examination.}
\label{tab:related}
\begin{adjustbox}{max width=\textwidth}
\begin{tabular}{>{\raggedright\arraybackslash}p{3.5cm}>{\raggedright\arraybackslash}p{3.0cm}>{\raggedright\arraybackslash}p{3.3cm}>{\raggedright\arraybackslash}p{4.4cm}}
\toprule
System & Questions from the student's submission & Adaptive examination by the system & Deployment reported \\
\midrule
\citet{lehtinen2023questions} & Yes, code & No: multiple choice, answered once & One exercise in a live course, 324 students consenting, 291 analyzed \\
\addlinespace
\citet{canico2025integrating} & Yes, code & No: single choice, optional & Preliminary: 13 students, 889 answers \\
\addlinespace
\citet{goodfellow2026questions} & Yes, code & No: multiple choice, formative & Introductory course of 239 students, 134 generated questions \\
\addlinespace
\citet{ipeirotis2026} & No: rubric and project description & Yes, voice & Replacing a final examination, cohorts of 36 and 37 \\
\addlinespace
Socratic Mind \citep{hung2024socratic,lee2026socratic} & No: instructor questions and materials & Yes, text and voice & One optional extra-credit assignment: 579 assigned, 147 never attempting, 171 analyzed \\
\addlinespace
TAMIGO \citep{anishka2024tamigo} & No: course materials & No: administered by teaching assistants & 411 students in 151 groups \\
\addlinespace
\citet{church2025viva} & Yes, essay & Yes & Proof of concept, no participants \\
\addlinespace
AVERT \citep{vintila2024avert} & Yes, code & Yes & Work in progress, no participants \\
\addlinespace
\citet{frankford2026chatbot} & Yes, code (by design) & Yes (by design) & Framework with a proof-of-concept prototype, no deployment \\
\addlinespace
ATC (this paper) & Yes, code & Yes, text & Grade of record per assignment: 247 students, 546 sittings, two examination languages \\
\bottomrule
\end{tabular}
\end{adjustbox}
\end{table}

The idea of asking students automatically generated questions about their own programs was introduced by \citet{lehtinen2021ask}. \citet{lehtinen2023questions} generate multiple-choice questions from the submitted Python program by static analysis, citing literal line numbers and drawing distractors from the student's own identifiers. Caniço and Santos \citeyearpar{canico2025integrating} instantiate templates with the submission's structure, identifiers and execution data, so that a question names the student's own functions and variables, and describe their deployment as preliminary. \citet{goodfellow2026questions} use generative AI to produce multiple-choice questions from each student's submitted code and restrict them to formative use. Anchoring a question to the student's own code, and to a specific line of it, is therefore not new. What these systems lack is an examination: a sequence of questions that responds to the student's answers and ends in a grade.

Another line conducts the examination but does not start from the student's submission. Ipeirotis and Rizakos \citeyearpar{ipeirotis2026} deploy a voice agent that examines students from a rubric and a text description of the project, as a replacement for the final examination. Socratic Mind \citep{hung2024socratic} opens with instructor-defined questions and adapts its follow-ups to the student's answers. Instructors configure the prompts, the desired answers and the rubrics, and a feature drafts questions from lecture materials. Where code is discussed, the instructor supplies it. A later study of the same system \citep{lee2026socratic} reports an optional extra-credit assignment, formative rather than an assessment of record, in a mixed text and voice interface. Autonomous adaptive examination of a real cohort is therefore not new either. What these systems do not do is read what the student handed in.

How the examination is graded differs too. Ipeirotis and Rizakos \citeyearpar{ipeirotis2026} use a panel of language models that score the same material, followed by a consultation step between them, a method also proposed for evaluation beyond education \citep{verga2024juries}. Agreement among their models was poor before that step, and after it about 60\% of ratings fell within one point of one another and 29\% matched exactly. Using a language model as a judge of open-ended text is common practice \citep{zheng2023judge,chiang2023alternative}, and it has been reviewed for educational assessment specifically \citep{emirtekin2025assessment}. ATC instead controls what each judge is allowed to see. The Code Reviewer reads the code and not the examination, the Grader reads the examination without direct access to the code, and their scores are combined by fixed arithmetic, with understanding and authorship kept as separate axes.

Three published designs have both properties. \citet{church2025viva} describe a viva voce simulator that questions a student adaptively on their own submitted work, with a system prompt that directs the first question to a specific non-trivial detail of the submission. The submission there is written prose rather than source code, and the paper is a proof of concept, with one demonstration transcript on an undergraduate essay and a judgment passed to a human examiner rather than recorded as a grade. Vintila's AVERT \citep{vintila2024avert} has both properties for source code: a chatbot questions students on their submitted code about how they developed it, adapting its questions to their answers. It too is a proof of concept, presented as work in progress with an evaluation planned. \citet{frankford2026chatbot} review conversational assessment in programming education and propose a hybrid framework for automated assessment systems, in which a code analysis engine extracts facts from the submission and a conversational agent builds its questions from them and adapts to the student's answers. Its authors present it as a design synthesis, with a proof-of-concept prototype and no classroom deployment. TAMIGO \citep{anishka2024tamigo} comes close without having both properties: it generates oral examination questions from course materials and assignment descriptions, and teaching assistants administer them, so the system never questions a student. These three designs describe the conjunction of the two properties, but to our knowledge none has been used to examine students in a course.

The gap, then, is that the two properties have not been brought together in a running course. ATC is a deployed instance of that conjunction, producing a grade of record per assignment across a term on a full undergraduate cohort. None of its parts is claimed as new: the anchoring belongs to the line on questions about learners' code, the autonomous adaptive dialogue to Ipeirotis and Rizakos and to Socratic Mind, and the architecture that joins them to Church et al., to Vintila, and to Frankford et al. What was missing was evidence about what such a system does when it is run on a real course at scale, which Section~\ref{sec:observed} reports.

\section{ATC system design}\label{sec:design}

\subsection{Design goals}\label{sec:goals}

The system was built to answer one question at the scale of a full course: does this student understand the code they submitted? Three commitments follow from that question.

The first is that questions must be anchored in the student's own submission rather than in course material or a description of the work. A question that can be answered from the assignment brief, from generic operating systems theory, or from an explanation of the code the student read shortly before the exam does not separate understanding from recall. The second is that the assessment must not require a human examiner in the loop, because a human oral defense for every student on every assignment is the cost that made oral assessment rare \citep{ellis2026scale}. The third is that the resulting grade must be defensible after the fact. This means structured output, a full audit trail, and a grading path whose arithmetic a student or an appeals committee can follow step by step, rather than a single holistic judgment by a model.

One further commitment is that the grade is not meant to depend on whether AI was used. The system records an advisory signal of machine-generated style for staff (Section~\ref{sec:authorship}), which never enters the arithmetic that forms the grade and is not validated here as a detector. Students were permitted to use AI assistants, and a student who used one and can explain every decision scores well.

\begin{figure}[t]
\centering
\includegraphics[width=\linewidth]{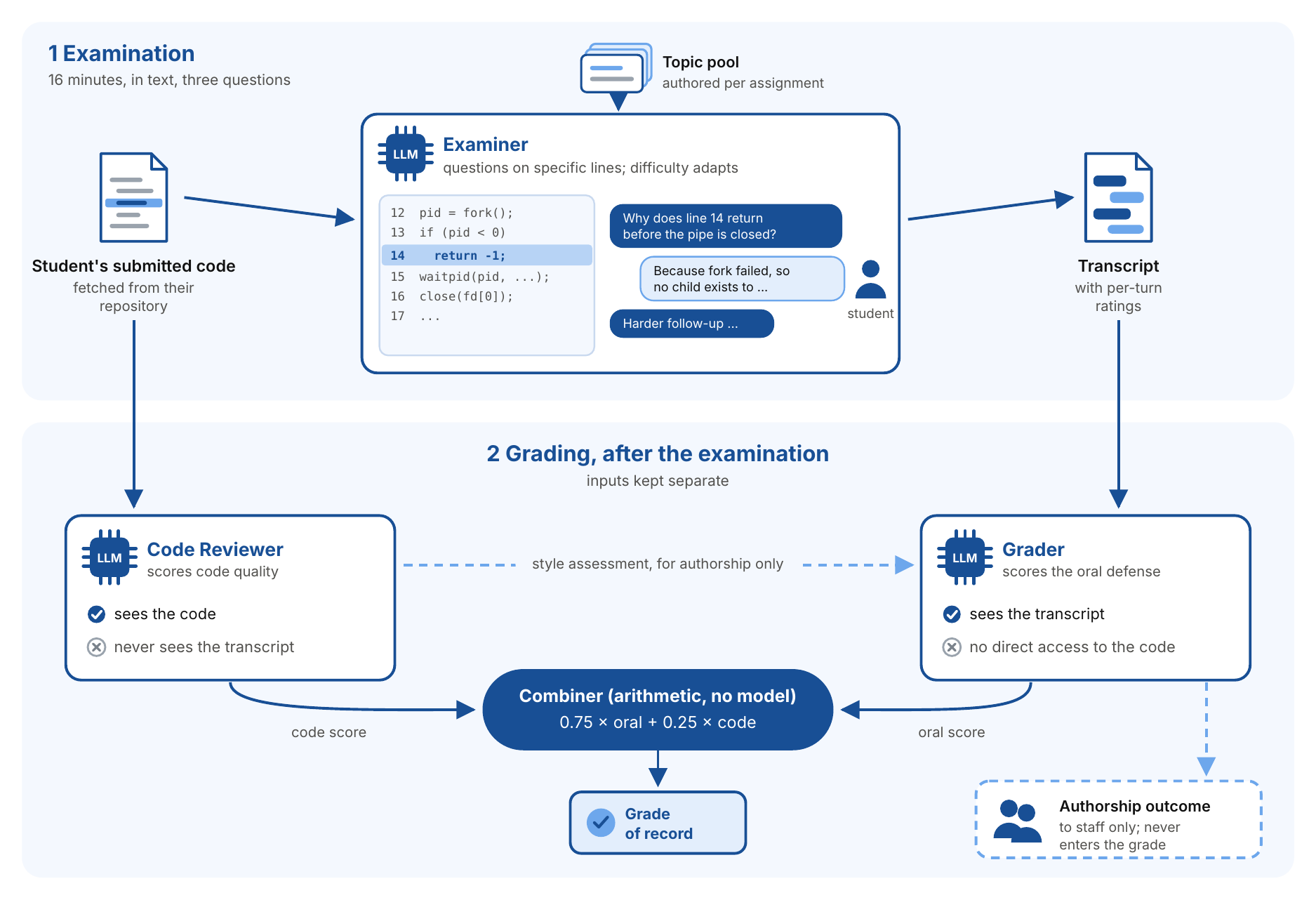}
\caption{Examination and grading. The examiner questions the student on their own submitted code. Grading then runs in three steps: a Code Reviewer that never sees the examination, a Grader that scores the transcript, and a fixed arithmetic combination of their two scores, 0.75 times the oral score plus 0.25 times the code quality score. The Grader has no direct access to the code. It sees code only where the examiner quoted it, together with the scores the examiner recorded for each answer. The Code Reviewer's style assessment reaches the Grader for the authorship decision, and the authorship outcome goes to staff without entering the grade.}
\label{fig:pipeline}
\end{figure}

This section describes the system as configured for the deployment reported in Section~\ref{sec:deployment}. Figure~\ref{fig:pipeline} summarizes how an examination runs and how it is graded, and \ref{app:implementation} gives the implementation detail an adopter would need.

\subsection{Intake and code context}\label{sec:intake-design}

Each assignment specifies a program to be built. Students receive six to seven pages of instructions, including a file tree and skeleton code, extend the skeleton, and hand in every file the instructions require. Submission is through GitHub Classroom, one private repository per student per assignment, and at exam time the system fetches the submitted files directly from the student's repository. The fetched files are assembled into a code context that is given to the examiner and displayed to the student.

\subsection{Code-anchored adaptive questioning}\label{sec:questioning}

\begin{figure}[!t]
\centering
\includegraphics[width=\linewidth]{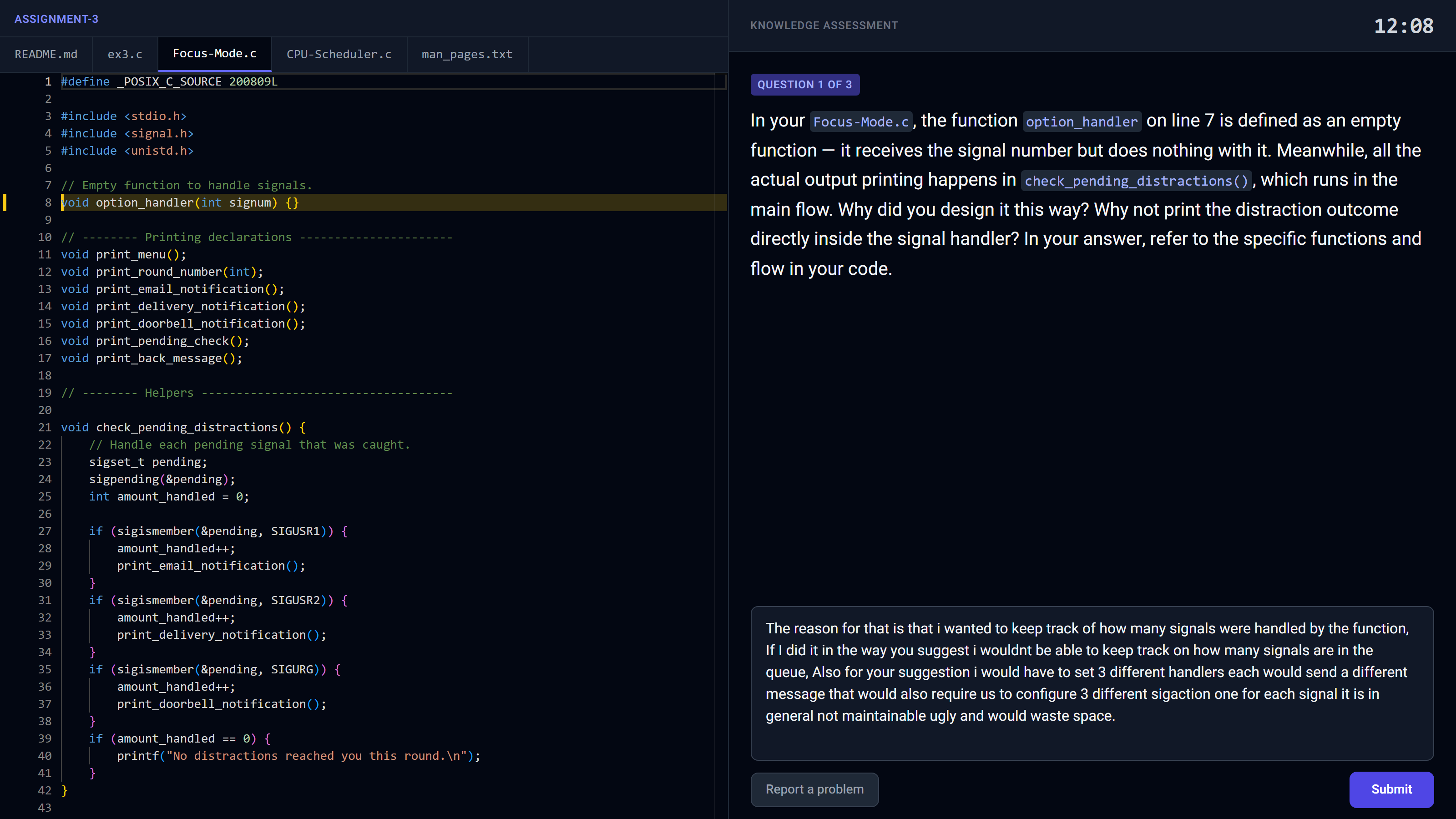}
\caption{The examination screen at the first question of an exercise 3 examination held in English, chosen among the English sittings on that exercise as one close to the exercise median. The student had typed an answer to the first question and was about to submit it, with 12 minutes 8 seconds remaining. The submitted code is on the left, scrolled to the line the examiner asked about, which is highlighted. The question, the answer box and the clock are on the right. The screenshot was produced after the deployment by replaying the stored sitting through the interface as deployed, rather than captured during the examination. Figure~\ref{fig:example-dialogue} gives the three questions and answers.}
\label{fig:example}
\end{figure}

Each exam consists of three questions and allows 960 seconds (16 minutes), extended to 1140 seconds (19 minutes) for students entitled to extra time, as described in Section~\ref{sec:safety}. The clock stops while the examiner generates the next question, so the time a student waits for the model does not count against the allowance. Questions are asked and answered in text. The interface is a split screen with the submitted code on one side and the dialogue on the other (Figure~\ref{fig:example}). Each question together with the student's answer to it is called a turn.

Questions are generated during the examination from topics. Each assignment has a topic pool, built before the assignment opens. Once per course, staff set the conceptual dimensions that suit the subject. For operating systems these are design choice, edge case, error handling, interface depth, code flow, and cross-file interaction. For each assignment, the system then generates topics automatically from the assignment, and the course staff remove those that are unfair at that point in the semester, that the assignment did not require, that can be answered from the instructions alone, or that cannot be answered in about five minutes.

A topic is the basis for a question and contains no question text. It specifies the files it targets, a focus, one of the dimensions, a difficulty label, a note telling the examiner what to probe, and the topics it must not be combined with in one exam. One topic, for example, targets \texttt{tournament.c}, with the focus ``the \texttt{fork()} call that creates four child processes'', the dimension error handling and the difficulty medium. Its note asks the examiner to have the student show where the processes are split, and to check whether the code tests the return value of \texttt{fork()} for failure. Which topics are offered is decided by deterministic code with no language model, which spreads questions across files and dimensions and excludes repeats and conflicting topics. The examiner receives the student's code together with a small menu of these topics, and a topic becomes a question only when the examiner meets it with that code: the examiner formulates the question from the code itself, quoting the student's identifiers and referring to specific lines. The same topic can therefore take many forms, even on the same code. The examiner can also direct the interface to scroll to and highlight a specific file and line, so the student can see the code being asked about. It may also change the conditions of a question, asking what the code would do with a different input, or propose a change to the code and ask how its behavior would differ. If the examiner notices a flaw in the code, it may set the planned topic aside and ask the student to explain that flaw instead. The exact wording of a question therefore cannot be known in advance, and a student prepares by studying their own code. Figure~\ref{fig:example-dialogue} shows an example exchange.

\begin{figure}[p]
\centering
\makebox[\linewidth][c]{\begin{minipage}{1.14\linewidth}
\footnotesize
\raggedright
\textbf{Question 1} (medium, highlighted \texttt{void option\_handler(int signum) \{\}} in \texttt{Focus-Mode.c}).
\emph{In your \texttt{Focus-Mode.c}, the function \texttt{option\_handler} on line 7 is defined as an empty function --- it receives the signal number but does nothing with it. Meanwhile, all the actual output printing happens in \texttt{check\_pending\_distractions()}, which runs in the main flow. [\ldots] Why not print the distraction outcome directly inside the signal handler? [\ldots]}

\smallskip
\textbf{Answer.} ``The reason for that is that i wanted to keep track of how many signals were handled by the function, If I did it in the way you suggest i wouldnt be able to keep track on how many signals are in the queue, [\ldots] it is in general not maintainable ugly and would waste space.''

\smallskip
\textbf{Examiner's evaluation, 3 of 5.} Valid practical reasons grounded in the code, but the operating systems reason for keeping a handler minimal is missing: \texttt{printf} is not safe to call from a signal handler. A score of 3 holds the next question at medium.

\medskip
\textbf{Question 2} (medium, highlighted \texttt{void reset\_data(Process processes[], int processes\_count) \{} in \texttt{CPU-Scheduler.c}).
\emph{Looking at your \texttt{CPU-Scheduler.c}, your \texttt{execute\_and\_report} function calls \texttt{reset\_data} at the end of each algorithm run before the next one begins. What specific state does \texttt{reset\_data} clear, and why is it necessary to reset that state between algorithm runs? [\ldots]}

\smallskip
\textbf{Answer.} ``The state that is cleared is: The remaining burst time [\ldots] The waiting time needs to be reset for a new algorithm run since the algorithm starts over for each algorithm run. The pid is set to -1 to make sure that we dont accidently try to access a process that exited. If i skipped this step, The algorithms that run sequencially may not run at all [\ldots]''

\smallskip
\textbf{Examiner's evaluation, 4 of 5.} The answer names all three fields and what goes wrong without the reset. A score of 4 steps the next question up to hard.

\medskip
\textbf{Question 3} (hard, highlighted \texttt{void check\_pending\_distractions() \{} in \texttt{Focus-Mode.c}).
\emph{In your \texttt{Focus-Mode.c}, at the end of each round you call \texttt{check\_pending\_distractions()} while signals are still blocked, and only afterward you call \texttt{sigprocmask(SIG\_UNBLOCK, \&block\_mask, NULL)}. [\ldots] POSIX does not guarantee any particular delivery order when multiple standard signals are unblocked simultaneously. How does your code guarantee that the output always appears in the correct order [\ldots] regardless of which signals are pending? [\ldots]}

\smallskip
\textbf{Answer.} ``The function sigismember checks if a specific signal is part of the set, By checking if the signals are part of the set in the order 1,2,3 we make sure that if a signal is in the set it is by order with the control flow of the function. Signals are saved only once even if the same one was sent multiple times. [\ldots] The order of arrival doesnt matter.''

\smallskip
\textbf{Examiner's evaluation, 5 of 5.} The output order comes from inspecting the pending set in a fixed sequence while the signals are still blocked, rather than from the order of delivery, and the student also noted that a standard signal is recorded once however often it is sent.
\end{minipage}}
\caption{The exchange in the examination shown in Figure~\ref{fig:example}: the three questions, the student's answers and a summary of the examiner's recorded evaluation of each. Questions and answers are verbatim, including the student's spelling, and [\ldots] marks text omitted for length. The Grader confirmed the examiner's three scores. Under the scoring table described in Section~\ref{sec:grading}, their average of 4.0, with at least one hard question, maps to the range 90 to 96, and the Grader placed the sitting at 91, against an exercise median of 88.}
\label{fig:example-dialogue}
\end{figure}

Difficulty adapts within the exam, over three levels: easy, medium and hard. The first question is chosen between two topics of medium difficulty. After each answer the examiner records an understanding score from 1 to 5. For each of the two later questions it then receives two topics at each level and chooses the level it judges the next question should have, guided by a stepping rule: a score of at least 4 steps up one level, a score of at most 2.5 steps down one level, and intermediate scores hold the current level. In Figure~\ref{fig:example-dialogue}, a score of 3 held the second question at medium, and a score of 4 stepped the third up to hard. The selector is seeded separately for each sitting, so the topics offered vary across students and each exam can be reproduced from its stored state (\ref{app:implementation}).

Two conversational behaviors complete the questioning loop. When a student does not attempt a question (empty or garbled input, a request for a hint, or an answer to something else), the examiner may ask once more for an answer, and a guard prevents this from repeating within a question. Any attempt, including a wrong one, is graded rather than re-asked. Separately, a student may ask once per exam for a different question. The first request is granted with a topic at the same difficulty, and any further request is declined. Both the original reply and the retry, and the exact wording of any switch request, are preserved in the transcript for the Grader and for later review.

\subsection{Three-agent grading with separated inputs}\label{sec:grading}

Grading involves the three agents introduced in Section~\ref{sec:introduction}. The examiner records a score for each answer during the exam (Section~\ref{sec:questioning}). After the exam ends, the Code Reviewer and the Grader work from separated inputs, and their scores are then combined by arithmetic.

The Code Reviewer sees the assignment description and the submitted source code and has no access to the exam transcript. It produces the code quality score, feedback for the student, notes for staff, and a structured assessment of how machine-generated the code's style looks, which is discussed in Section~\ref{sec:authorship}. The Grader sees the assignment description and the full transcript, together with that style assessment, and has no direct access to the submitted code. It re-evaluates the understanding scores the examiner recorded for each turn and may override them. It then produces the oral score from two quantities, the average of those scores and the difficulty of the questions the student answered, as described below, and writes both student feedback and a staff report. The combination step is arithmetic: the final grade is 0.75 times the oral score plus 0.25 times the code quality score, rounded to the nearest integer, with no caps and no penalties applied to it. Sittings that reach the time limit are graded on the partial transcript, with the Grader told which questions were answered and which were never reached.

The separation is central to the design. If one agent saw both the code and the transcript, clean code would flatter the oral score and a fluent defense would flatter the code quality score, and the two numbers we most want to compare would be formed from the same evidence. Even so, three routes carry code-derived information to the Grader: the lines the examiner quoted, the per-turn scores recorded by an examiner that had read the code, and the style assessment. Section~\ref{sec:lessons} considers what they imply. Keeping the two inputs separate is what makes the result in Section~\ref{sec:observed} meaningful, and it is also why the two scores are combined by a formula rather than by a further model judgment.

\paragraph{Scoring rules and models}
On the examiner's understanding scale, a 5 is an answer that explains both what the student's code does and why, in terms of the underlying operating systems mechanism and specific to that code. A 3 is an answer that describes what the code does without explaining why. A 1 is an answer that cannot explain the code or contradicts it. Forgetting an exact string or a variable name, or answering correctly at a higher level of abstraction, does not bring a score below 4. The Grader then re-scores every answer from the transcript on the same scale, with its own rules and with the instruction to favor the student where an answer is ambiguous and to grade the answer the student gave rather than the fuller answer the model might have preferred. A short answer that is correct scores at least 3, and scores of 2 or below are reserved for answers that are wrong or incoherent.

The Grader forms the oral score in two steps. In the first, a fixed table gives a range of oral scores from two inputs: the Grader's understanding scores for the student's answers, divided into seven bands, and the difficulty profile of the questions the student answered, one of three: at least one hard question, medium questions only, or at least one easy question. All but one of the ranges are five to fourteen points wide, the exception being the lowest band with an easy question, which runs from 0 to 38. Within a band the three ranges overlap, and a harder profile receives a higher range. Between bands, ranges overlap down to a mean of 2.5 and touch or leave a gap below it. In the second step, the Grader places the student within that range by its judgment of the answers, except when fewer than three answers could be graded, in which case it places the student at the bottom of the range. Figure~\ref{fig:example-dialogue} gives an example.

The Code Reviewer assigns a score from a table of six tiers. Each tier is described jointly by coverage, meaning which required files and features are present, and by correctness, meaning how far the code that is present implements the specification, judged by reading it, including whether system call return values are checked and memory is freed. A missing required file costs between 10 and 15 points, and where a submission falls between two tiers the higher one is taken.

The agents ran on Claude Opus 4.6, called through the provider's application programming interface, with their output forced into a fixed schema (\ref{app:implementation}). When Opus returned no answer, a call fell back to Claude Sonnet 4.6 and then, as a last resort, to a Gemini model from Google, so that an exam in progress would continue instead of failing. The Grader and the Code Reviewer ran at temperature 0.1, and the examiner at a temperature drawn for each sitting between 0.2 and 0.4 and stored with it.

\subsection{Authorship as a separate and advisory axis}\label{sec:authorship}

Understanding and authorship are kept on separate axes. During the exam the examiner rates each answer for authorship confidence, that is, how confident it is that the student wrote the code (high, medium, or low), independently of whether the answer was correct. The question behind this rating is whether the student speaks about the code as its author, referring to implementation details that only someone who worked on it would know, as opposed to reciting general theory that fits any submission. A student who defends their own design choice wrongly but with such details therefore receives a low understanding score but is still rated as the likely author. This per-turn rating is the primary signal.

A second, supporting signal is the Code Reviewer's style assessment, which looks for stylistic markers of machine-generated code and returns a level from 0 to 3 with an explanation. It cannot by itself produce an adverse authorship finding.

The two signals are combined through a fixed table into one of three outcomes: authorship established, partially established, or not established. After the exam the Grader can also raise a flag, called an integrity flag in the system, when a student consistently fails to explain their own code. Despite its name, the flag concerns understanding rather than authorship, and the same rule applied in every sitting. The authorship outcome and the flag are shown to staff for human review, and neither enters the grade. An automated pipeline should not convict a student of misconduct, so concerns are routed to people.

Two staff reports from the deployment show the two axes apart. In the first, the Code Reviewer found strong signs of machine-generated code, the same pattern repeated across four files with no traces of development. In the examination, however, the student explained why standard input is a stream that can be read only once and walked through a six-stage shell pipeline step by step. The system gave an oral score of 92 and recorded authorship as established. In the second, the answers themselves pointed away from authorship: the student could not explain a \texttt{break} statement in a loop of their own script, claimed that the loop kept running, contrary to the submitted code, and then asked to switch the question, saying they had not written that line. The system gave an oral score of 69, recorded authorship as not established, and passed the case to staff.

\subsection{Safety and fairness rules}\label{sec:safety}

Emotional distress is handled as an expected case. The system has no view of the student beyond the text they type, so what it acts on is what the student writes: a message stating or plainly conveying that they are panicking, that they are crying, or that they cannot continue. On such a message the examiner ends the exam immediately and gently, the session is marked accordingly, and no grade is computed. A teaching assistant is notified with the triggering message and a link to the transcript, and the student is exempted from the lock that otherwise prevents a further attempt without staff action (Section~\ref{sec:booking}), so that a fresh session can be arranged. The trigger criteria are strict, so that the ordinary frustration of an examination does not end a graded sitting: frustration, tiredness, saying ``I don't know'', or calling a question unfair do not qualify.

Adversarial input is handled without changing the examiner's tone. Instructions planted in an answer to redirect the examiner are ignored in the dialogue, flagged in the structured output, and reported to the Grader and to staff, and the flag alone does not change the score. Hostile or coercive demands are declined in a scripted, even-handed way and recorded verbatim.

Students who believe the displayed code is not theirs can say so during the exam. The complaint is recorded with full session metadata, staff are emailed, the exam ends on an acknowledgment screen, and the account is locked so that no new exam can start until a human has reviewed the case, with the transcript preserved for that review. Section~\ref{sec:operation} reports how these provisions behaved in testing and during the deployment.

\label{sec:languages}Two further provisions concern access rather than conduct. Exams run in either of two languages, selected by the student before the session begins: Hebrew, the language of instruction, which is the default, or English. The examiner instructions are maintained separately for each language instead of being translated at run time, so that the examination assesses understanding of the code rather than command of academic English. Students entitled to extended time are identified automatically from a list supplied by the course, with no manual step per session, so they do not have to disclose the accommodation at the examination.

\section{Deployment and study design}\label{sec:deployment}

\subsection{Course and cohort}\label{sec:cohort}

The system was deployed in Introduction to Operating Systems, an undergraduate computer science course at Bar-Ilan University, during the spring semester of 2026. The roster held 247 students. Each of the three exercises was followed by an examination. A sitting is one student's attempt at one examination, and 546 sittings were recorded in total.

A student's record on an exercise is called usable when the student submitted the exercise and sat its examination, and the sitting was not ended on the distress path (Section~\ref{sec:safety}). Section~\ref{sec:analysis} states this exclusion rule in full. The cohort is not a single degree track: among the 216 students with a usable exercise 3 record, on whom the primary criteria are evaluated, 123 were on the pure computer science track, 33 on computer science with mathematics, 24 on computational biology, 23 on software engineering, and 13 on other tracks. Programs with fewer than 20 of these students were grouped as other. Degree grade point average was available for 214 of the 216, and exercise 3 analyses pairing an examination score against grade point average run on those 214.

\subsection{The crossover design, and what varied}\label{sec:crossover}

Two lecture groups were opened for the course, and students registered to whichever they chose, forming the groups denoted S with 84 students and L with 163. The groups sat the same exercises on the same schedule. On exercises 1 and 2 they differed in what they were told in advance about how their work would be graded. On exercise 1, group S was told that the exercise would be graded by the AI examination system, and group L that it would be graded by the course's automatic tests, which run the submitted program against fixed test cases (Section~\ref{sec:exercises}). On exercise 2 the two groups swapped, and on exercise 3 the whole cohort was told it would be graded by the AI examination system. We call the two conditions arms, although the groups were not randomized. We name them told-AI and told-autograde, because the distinction concerns what the student knew when writing the code rather than which score was computed. Table~\ref{tab:crossover} summarizes the design.

\begin{table}[t]
\centering
\small
\caption{What each group was told, before submitting, about how each exercise would be graded: by the AI examination system (told-AI) or by the course's automatic tests (told-autograde).}
\label{tab:crossover}
\begin{tabular}{lccc}
\toprule
 & Exercise 1 & Exercise 2 & Exercise 3 \\
\midrule
Group S (84 students) & told-AI & told-autograde & told-AI \\
Group L (163 students) & told-autograde & told-AI & told-AI \\
\bottomrule
\end{tabular}
\end{table}

The order followed a gradual rollout, chosen to reduce the risk of a first deployment of a new system: the oral examination was introduced for one group on exercise 1, moved to the other group on exercise 2, and used for the whole cohort on exercise 3. Because the order was reversed between the groups, every student was assigned one told-AI exercise and one told-autograde exercise on exercises 1 and 2. This supports a comparison, on each exercise, of how the two modes were received and of how automated test scores behaved under each incentive. It does not support a comparison of learning or of grade quality, because mode is tied to exercise content within each group, and the groups differ in size and in track composition. Section~\ref{sec:observed} reports the reception and incentive comparisons on that basis and reports nothing about relative attainment across modes.

\subsection{Participation}\label{sec:participation}

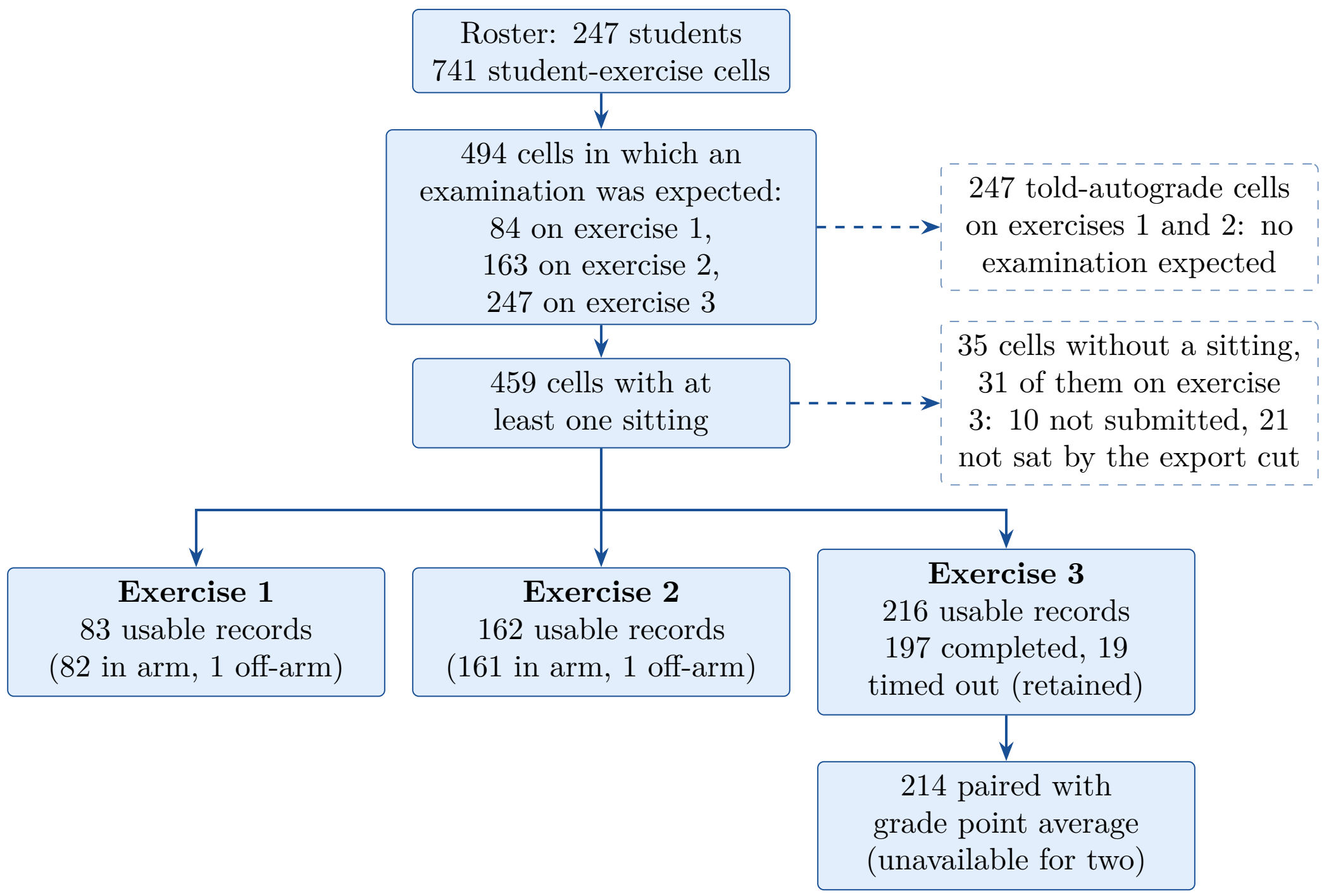
\begin{figure}[t]
\centering
\begin{adjustbox}{max width=\linewidth}
\begin{tikzpicture}[
  font=\footnotesize,
  box/.style={draw=atcdark, fill=atctint, rounded corners=2pt, align=center, text width=4.0cm, inner sep=4pt, execute at begin node={\hyphenpenalty=10000\relax}},
  side/.style={draw=atcdark!60, dashed, rounded corners=2pt, align=center, text width=4.0cm, inner sep=4pt, execute at begin node={\hyphenpenalty=10000\relax}},
  arr/.style={-{Stealth[length=2.2mm]}, thick, draw=atcdark},
]
\node[box] (roster) at (4.2,0) {Roster: 247 students\\741 student-exercise cells};
\node[box, text width=4.6cm] (exp) at (4.2,-2.0) {494 cells in which an examination was expected:\\84 on exercise 1, 163 on exercise 2,\\247 on exercise 3};
\node[side] (notexp) at (10.2,-2.0) {247 told-autograde cells on exercises 1 and 2: no examination expected};
\node[box] (sat) at (4.2,-4.0) {459 cells with at least one sitting};
\node[side] (short) at (10.2,-4.0) {35 cells without a sitting, 31 of them on exercise 3: 10 not submitted, 21 not sat by the export cut};
\node[box] (e1) at (-0.4,-6.6) {\textbf{Exercise 1}\\83 usable records\\(82 in arm, 1 off-arm)};
\node[box] (e2) at (4.2,-6.6) {\textbf{Exercise 2}\\162 usable records\\(161 in arm, 1 off-arm)};
\node[box] (e3) at (8.8,-6.6) {\textbf{Exercise 3}\\216 usable records\\197 completed, 19 timed out (retained)};
\node[box] (gpa) at (8.8,-8.8) {214 paired with grade point average (unavailable for two)};
\draw[arr] (roster) -- (exp);
\draw[arr, dashed] (exp) -- (notexp);
\draw[arr] (exp) -- (sat);
\draw[arr, dashed] (sat) -- (short);
\draw[arr] (sat.south) -- ++(0,-0.7) -| (e1.north);
\draw[arr] (sat.south) -- (e2.north);
\draw[arr] (sat.south) -- ++(0,-0.7) -| (e3.north);
\draw[arr] (e3) -- (gpa);
\end{tikzpicture}
\end{adjustbox}
\caption{Participation, counted in student-exercise cells (one student on one exercise). The usable records on exercises 1 and 2 each include one off-arm record, from a student who sat under the other group's arm. \ref{app:participation} gives the full accounting, including a third off-arm sitting that produced no grade.}
\label{fig:flow}
\end{figure}

We count participation in student-exercise cells, each cell being one student on one exercise. Under the crossover, a told-autograde student was not expected to sit an examination on that exercise, so the cells in which an examination was expected number 494: 84 on exercise 1, 163 on exercise 2 and 247 on exercise 3. Of those 494, 459 hold at least one sitting (Figure~\ref{fig:flow}). In the other 35, the student did not submit or did not sit. \ref{app:participation} gives the full accounting, including three sittings by students who sat under the other group's arm, two of which produced a grade. Every one of the 247 students on the roster sat at least one examination that produced a grade.

On exercise 3, to which the whole cohort was assigned, 10 of the 247 students did not submit and 21 submitted but did not sit, giving 31 excluded and 216 usable records. Of those 216, 197 were completed, and in 19 the time ran out before the last answer was submitted. These were graded on the partial transcript (Section~\ref{sec:timeouts}). Access to examination was never withdrawn, so the 21 who had not sat are counted as of the date on which the data were exported for analysis (the export cut) rather than as a final outcome. On the other two exercises the usable records number 83 on exercise 1 and 162 on exercise 2, each including one student who sat under the other group's arm.

\subsection{The three exercises and their automated tests}\label{sec:exercises}

Exercise 1 asked students to implement hooks for Claude Code in bash, exercise 2 to synchronize processes and threads in C, and exercise 3 to handle signals and schedule processes in C. Across the three exercises, staff removed about 15\% of the automatically generated topics (Section~\ref{sec:questioning}), and each exercise drew on between 20 and 43 distinct topics. Exercise 1 was scored by a shell-hooks tester and exercise 2 by a unit-test autograder for C, both on a scale of 0 to 100. Each awards points for exact output matches across a set of test cases, so a score is in effect a count of exactly matching cases, and a zero can mean correct logic with non-conforming printing. Both were run on the submissions of both arms, but their score counted as the exercise grade only in the told-autograde arm. Exercise 3 had no automatic grader.

\subsection{Procedure}\label{sec:procedure}

\paragraph{Submission intake}\label{sec:intake}
The system fetched each submission from the student's repository at examination time (Section~\ref{sec:intake-design}), so there was no separate upload step. Students could not change their repositories after the submission deadline, which ensured that the code examined was the code they had submitted.

\paragraph{Booking, examination periods and retakes}\label{sec:booking}
After submitting, each student booked an examination slot through an online calendar, and a reminder was sent automatically the day before the slot.

Each exercise opened with a booking window of three weeks, announced to students as the period within which they were expected to sit. The window was not enforced as a deadline. A student who had not booked within it, or who had booked and not attended, could sit afterwards on request, and every such request was granted by a member of the teaching staff. On every exercise, sittings arising from extended absences were still taking place well after the window had closed.

A per-account lock prevented a student from starting a further sitting on the same exercise without staff action. All students were told that a further attempt could be requested, and the grade of the last sitting counted for the course. A student seeking a further attempt asked the teaching assistant, who authorized it in the system and reports granting every such request. In practice this happened in two situations: after a technical problem during the examination, which leaves no grade, or when the student was dissatisfied with the grade received. Staff usually permitted the further attempt by invalidating the earlier sitting, whose record was retained rather than deleted. Where a student sat more than one sitting on the same exercise, the analyses that test the success criteria use the sitting that first produced a grade, and the later sittings are reported descriptively in Section~\ref{sec:retakes}.

\paragraph{The examination environment}\label{sec:room}
Examinations took place in person, in a dedicated room in the computer science building, and the room was unattended: no invigilator was present at any point, and the door was left unlocked. Each student sat at a workstation running in locked-down kiosk mode, which blocked external applications and browser access, and left personal belongings in a designated corner of the room. Students had the teaching assistant's email address and telephone number for any problem during a sitting. Two cameras recorded continuously, and the course instructor reviewed the recordings by sampling, partly live and partly on the following day. Sampling covered on average five minutes of each hour of recording, viewed at double speed, and checked that the student was alone, that belongings were in the designated corner, and that no phone or personal computer was in use. The cameras' built-in person detection, which runs on the device, tagged the intervals in which a person was present, so the sampled minutes fell on occupied time. Beyond that tagging, the footage received no automated analysis. An hour of recording spans roughly three sittings, so no sitting was reviewed as a matter of course, but every sitting had a chance of being observed. The review found no violation in any sitting it covered.

Unlike the self-recorded setting reported by \citet{ipeirotis2026}, in which students sat the examination on their own machines wherever they chose, here the machine and the room were controlled. In both cases the sitting was recorded and no member of staff was in the room. The room is a property of this deployment rather than of the software. A course adopting the system without a controlled room would obtain the same questioning and grading behavior, with weaker guarantees about the conditions in which answers were produced.

\paragraph{Appeals}\label{sec:appeals}
Every grade could be contested before a human. A student who disputed a grade could appeal by email within a week of receiving it, stating the grounds, and staff sent the transcript of the sitting on request. Appeals were decided by the teaching assistants after the examination period, by reading the transcript and the code. An accepted appeal corrected the course grade rather than the stored result, and every figure in this paper is the system's own grade before appeal. We analyze the stored grades because the object of study is what the system produced and because an appeal adds a human judgment only for the students who chose to appeal. Section~\ref{sec:operation} reports how many appeals were filed and how they were decided.

\paragraph{Surveys}\label{sec:surveys}
Before each exercise was graded, students received an email telling them how the exercise would be graded (by the AI examination system or by the course's automatic tests), with a link to a short course-evaluation form about that mode. After each AI examination, a second form was sent with the grade and the written feedback, because several of its items ask about the grade and the feedback received. The surveys were mandatory, but completion could not be enforced. \ref{app:survey} gives the three forms, and Section~\ref{sec:reception} reports the results.

\subsection{Consent, ethics and data handling}\label{sec:ethics}

The examinations and surveys were part of the course's regular assessment and evaluation. Students were told in advance that an AI model would examine them and that the examination room was recorded. The surveys were designed to improve the system and to report on its use to the university rather than as research instruments. The analyses reported in this paper were conducted on data generated by that assessment, with no additional research procedure imposed on students. Research use of the existing data was approved retrospectively by the Human Subjects Institutional Review Board of Bar-Ilan University (approval number 0109261257, 1 September 2026), which waived individual consent because the data were anonymized before analysis. The approval covers the examination records, the survey responses, degree grade point average and the course grades used for comparison in Section~\ref{sec:benchmark}. The camera recordings described in Section~\ref{sec:room} were not used as research data.

Degree grade point average and the course grades were matched to the course data by student identifier immediately before anonymization. The identified version was used only for that matching and was not retained. All analysis was carried out on the anonymized dataset, with session identifiers dropped and free-text survey responses excluded.

Under the model provider's published retention policy, inputs and outputs sent through its application programming interface (the agents' instructions, the assignment instructions, the files of the student's submission, the transcript and the Code Reviewer's style assessment) are deleted within 30 days by default \citep{anthropic_retention}, and under its commercial terms they are not used to train its models \citep{anthropic_commercial_terms}. The submitted code, the transcripts and the grading reports are kept in the deployment's own project on Google Cloud Platform (\ref{app:implementation}), which commits in its data processing terms to comply with the General Data Protection Regulation (GDPR) and provides integrated security controls \citep{google_cloud_gdpr}. The camera recordings were kept for one week and were accessible only to the course staff. No appeal drew on them, since the full text of every answer and complete logs were available.

\subsection{Success criteria and analysis}\label{sec:analysis}

Five success criteria, each with a numeric threshold, were fixed before any examination data had been extracted for analysis. Three concern the oral score and are evaluated on exercise 3, and two concern the automated test scores and are evaluated on exercises 1 and 2. Only these five are confirmatory. Every other coefficient in this paper is descriptive or exploratory, and Section~\ref{sec:observed} identifies which. Whether the system could run on the full course, and at what cost, is reported descriptively (Sections~\ref{sec:operation} and~\ref{sec:cost}).

The three criteria on the oral score apply the discriminant and convergent logic of construct validation \citep{campbell1959}. If the oral score reflects the student's account of their own code, it should not simply restate how good that code is, it should relate to how the student stands elsewhere in the degree, and it should relate to that more closely than to the quality of the code. The thresholds follow the conventional benchmarks for correlations, 0.10 for a small and 0.30 for a medium relationship \citep{cohen1988}.

\begin{itemize}
\item S1, discriminant: the upper bound of the interval for the correlation between the oral score and the code quality score is below 0.30, so that the data rule out even a medium positive relationship. With 216 students the interval extends about 0.13 on either side of the estimate, so an observed correlation above about 0.17 would fail S1.
\item S2, convergent: the lower bound of the interval for the correlation between the oral score and degree grade point average is above 0.10, so that the data rule out a negligible relationship. The threshold is low because grade point average is a coarse criterion (Section~\ref{sec:limitations}).
\item S3, separation, the primary criterion: the interval for the difference between these two correlations lies entirely above zero. This tests the central claim directly. The design predicts the direction, and a negative difference would contradict it.
\end{itemize}

The two criteria on the automated test scores concern the instrument the oral examination was meant to complement. The deployment rested on the premise that the course's automatic tests discriminate too little between students to carry a grade, and the crossover made it possible to ask whether students who expected automatic grading aimed their code at the tests.

\begin{itemize}
\item S4, premise: in the told-autograde arm, the automated test scores have a median at or above 90 and an interquartile width at or below 15, on exercises 1 and 2 separately, since their harnesses differ.
\item S5, incentive: on both exercises, the told-autograde median exceeds the told-AI median, with the interval for the difference excluding zero.
\end{itemize}

The surveys (Section~\ref{sec:surveys}) also carried ten hypotheses about student attitudes. They were written after the deployment and before any response was inspected, are not success criteria, and are described with their results in Section~\ref{sec:reception}.

\label{sec:sample}Exercise 3 is the primary analysis exercise for S1 to S3, because it is the only exercise on which both groups were assigned the AI examination. Almost every student entered it with a graded AI examination on exactly one earlier exercise (211 of the 216, while four had none and one had two), so prior exposure is close to balanced. S4 and S5 rest on the automated test scores of the told-autograde and told-AI arms of exercises 1 and 2 (Section~\ref{sec:ceiling}). The exclusion rule, fixed in advance, is the same on all three exercises and acts on recorded status. A record is excluded for non-submission, for not sitting, or for a sitting ended on the distress path, which leaves 83, 162 and 216 usable records on exercises 1 to 3 (Section~\ref{sec:participation}).\footnote{No sitting ended on the distress path.} Sittings in which time ran out are retained. Every usable record carries an oral score and a code quality score. Pairings with grade point average run on 81, 162 and 214 records for exercises 1, 2 and 3 respectively, because grade point average was unavailable for two students, both of whom sat exercises 1 and 3 (Section~\ref{sec:cohort}). The grade point average does not include the operating systems course itself, so the correlations between the oral score and grade point average need no correction for overlap between the two measures.

Rank correlations (Spearman) are the primary statistic throughout, because they are robust to single extreme values and to the bunching of scores near the top of their range, and product-moment correlations (Pearson) are reported in \ref{app:stats} as a robustness check. All intervals are bias-corrected and accelerated (BCa) bootstrap confidence intervals from 10,000 resamples of students, except the two for S5 (Section~\ref{sec:incentive}) and the interval for grading stability, which resamples sittings (Section~\ref{sec:reliability}). They are at the 95\% level unless a multiplicity adjustment is stated, which is the case only for the three primary reception hypotheses in Section~\ref{sec:reception}. The pairwise n is stated for every coefficient.

\section{Results}\label{sec:observed}

This section sets out what the deployment produced. Of the five success criteria (Section~\ref{sec:analysis}), four are met and one is not (Table~\ref{tab:criteria}). The section first describes how the system operated (Section~\ref{sec:operation}) and how the scores were distributed (Section~\ref{sec:distributions}). It then reports the criteria on the oral score (Section~\ref{sec:standing}) and on the automated test scores (Section~\ref{sec:ceiling}). The remaining subsections are exploratory or descriptive: robustness checks, the reliability of a single examination, a benchmark against course grades, how students evaluated the two modes, and the cost of the deployment. Together these results indicate that the examination ran at class scale and that its scores relate to other measures as the design predicts.

\begin{table}[t]
\centering
\small
\caption{The five success criteria, fixed before any examination data had been extracted for analysis (Section~\ref{sec:analysis}), and what was observed. Intervals are 95\% bootstrap intervals.}
\label{tab:criteria}
\begin{adjustbox}{max width=\textwidth}
\begin{tabular}{l>{\raggedright\arraybackslash}p{5.2cm}>{\raggedright\arraybackslash}p{4.2cm}l}
\toprule
Criterion & Requirement & Observed & Verdict \\
\midrule
S1, discriminant & Upper bound for the correlation of oral score and code quality score below 0.30 & $-0.073$ [$-0.205$, $+0.063$], $n=216$ & Met \\
S2, convergent & Lower bound for the correlation of oral score and grade point average above 0.10 & $+0.419$ [$+0.294$, $+0.523$], $n=214$ & Met \\
S3, separation & Interval for the difference between S2 and S1 lies entirely above zero & $+0.489$ [$+0.312$, $+0.662$], $n=214$ & Met \\
S4, premise & Told-autograde automated test median at or above 90, interquartile width at or below 15, on each exercise with an automatic grader (exercises 1 and 2) & Median 100, width 2, on both exercises & Met \\
S5, incentive & Told-autograde automated test median above told-AI median, interval excluding zero, on both exercises & Median difference 0.0 on both exercises & Not met \\
\bottomrule
\end{tabular}
\end{adjustbox}
\end{table}

\subsection{Operation in practice}\label{sec:operation}

The deployment ran on a single examination workstation. From the moment a student submitted an answer to the examiner's reply, the examiner took a median of 28 seconds (90th percentile 37 seconds, maximum 76 seconds, over 1,646 calls). Preparing the first question took a median of 22 seconds (546 sittings), for 2,192 examiner calls in all. The examination clock was stopped during these waits (Section~\ref{sec:questioning}). The examiner fell back to Sonnet in a single sitting, on all 3 of its calls, and no call reached Gemini. Of the 546 sittings, 43 were held in English, by 22 of the 247 students, and the rest in Hebrew.

The distress path, the prompt-injection flag, the handling of hostile demands and the flow for reporting code that is not the student's all worked when tested before the deployment, and none was triggered during it. Of the 83, 162 and 216 usable sittings on exercises 1 to 3, authorship was recorded as not established in 9, 7 and 3, as partially established in 3, 1 and 0, and as established in the rest. Staff looked at these outcomes, and none of them affected a grade.

Over the three exercises, 21 students filed 23 appeals against a grade the system had assigned, under the procedure described in Section~\ref{sec:appeals}. They came from the 461 student-exercise cells with a graded sitting (\ref{app:participation}). Eighteen contested only the code review, and five contested the oral examination, one alone and four together with the code review. Sixteen were upheld, two of them in part, and seven were rejected. All five that contested the oral examination were upheld. Ten appeals, on exercises 1 and 2, concerned code-review deductions and answers marked wrong. In one of them the answer had been garbled by a switch between Hebrew and English. The other 13, all on exercise 3, arose from one recurring error. A file that the course staff supplied with the assignment contained a fault, and in 148 of the 216 exercise 3 reviews the Code Reviewer deducted points for it. The course added 7 points to the code quality score of every affected student, whether or not they had appealed. The deduction was the system's error, since it failed to attribute the fault to the staff, so here too the analyses use the stored scores, as on exercises 1 and 2 (Section~\ref{sec:appeals}). Section~\ref{sec:robustness} repeats S1 and S3 on the corrected scores.

\paragraph{Timeouts}\label{sec:timeouts}
As in any timed examination, some students did not finish. The time ran out before the student had submitted the last answer in 7 of the 83 usable sittings on exercise 1, 21 of 162 on exercise 2 and 19 of 216 on exercise 3. In 45 of these 47 sittings the student was still typing an answer, the third in all but one case, and the typed text was captured and graded as an answer. In the other 2, nothing had been typed for the third question.

\begin{figure}[!t]
\centering
\includegraphics[width=\linewidth]{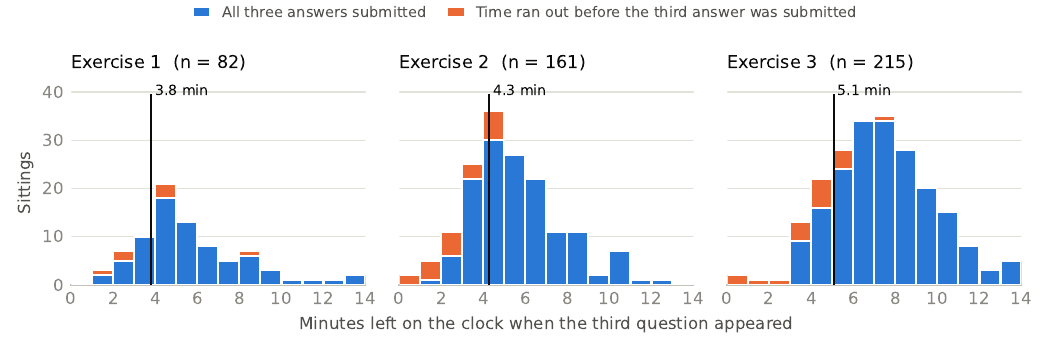}
\caption{Time left on the examination clock when the third question appeared, for the usable sittings of each exercise that reached it. Colors show whether the student submitted all three answers or the time ran out before the third answer was submitted. The vertical line marks the median time taken to answer the third question in completed sittings. The allowance was 16 minutes, or 19 with extra time. The last bin includes everything above 13 minutes.}
\label{fig:q3time}
\end{figure}

Most students had enough time for the third question (Figure~\ref{fig:q3time}). When it appeared, a median of 5.2, 5.0 and 7.2 minutes remained on exercises 1 to 3, and at least three minutes remained in 88, 89 and 98 percent of the sittings. A third answer in a completed sitting took a median of 3.8, 4.3 and 5.1 minutes. The students whose time ran out had spent longer on the first two questions, a median of 12.8 minutes against 9.8 in completed sittings, which left them a median of 3.8 minutes for the third. In 40 of the 44 who were still typing the third answer, the captured text ran to at least 20 words. Its median length was 53 words, against 73 for third answers that were submitted. The oral scores of students whose time ran out did not differ significantly from those of students who completed the examination on exercises 1 and 2, with medians of 80 against 80 and 80 against 81 (Mann--Whitney $p$ = 0.68 and 0.26). On exercise 3 their median was lower, 83 against 88 ($p$ = 0.07).

\paragraph{Sittings after the first}\label{sec:retakes}
Some students sat the same exercise more than once. Sittings carrying no final grade, at which the student reported a problem partway through and stopped, are excluded (14 of the 546), as is one confirmed technical fault. That leaves 531 sittings, of which 70 are retakes, in 58 student-exercise cells held by 54 of the 247 students. From the first to the last kept sitting in a cell, the median change in oral score is +11.0 points, with quartiles at +1.25 and +19.0. Of the 58 cells, 45 are higher on the last sitting, 12 lower and 1 tied. Part of this rise may reflect what students learned between sittings, but the comparison cannot separate learning from selection. A student dissatisfied with a grade could have the sitting invalidated and sit again, and the last sitting's grade counted for the course (Section~\ref{sec:booking}). The first sittings that a retake follows were therefore selected for being low, and a second sitting after a low score tends to come out higher even when nothing has changed, because part of the low score was chance (regression to the mean). Except where stated, the other analyses in this paper use the first sitting in each cell that produced a grade, so retakes do not enter them. \ref{app:participation} gives the full accounting.

\subsection{Distributions}\label{sec:distributions}

\begin{table}[t]
\centering
\small
\caption{Distributions of the oral score, the code quality score, the automated test score and degree grade point average, on the usable records. The mean of two orals averages a student's exercise 3 oral score with their oral score on exercise 1 or 2. It exists for the 212 students who have both, and it is the measure that the benchmark in Section~\ref{sec:benchmark} uses. The automated test rows cover the students who sat the oral examination on that exercise, so they describe the same submissions as the oral rows. Section~\ref{sec:ceiling} reports the full told-AI arms ($n = 84$ and 163). The exercise 3 code quality row gives the stored scores, which carry the 7-point deduction described in Section~\ref{sec:operation} on 148 of the 216 reviews. The deduction lowers the row's mean and its share at or above 90. On the corrected scores the mean is 86.02 and the standard deviation 6.92.}
\label{tab:distributions}
\begin{adjustbox}{max width=\textwidth}
\begin{tabular}{lccccccccc}
\toprule
 & n & mean & SD & median & IQR & min & max & at or above 90 & distinct values \\
\midrule
Exercise 1 oral & 83 & 76.96 & 12.55 & 80.0 & 17.0 & 47 & 100 & 12.0\% & 31 \\
Exercise 2 oral & 162 & 77.68 & 12.57 & 81.0 & 13.0 & 15 & 93 & 7.4\% & 33 \\
Exercise 3 oral & 216 & 86.22 & 8.79 & 88.0 & 11.0 & 48 & 97 & 43.5\% & 26 \\
Mean of two orals & 212 & 82.07 & 8.43 & 83.5 & 10.5 & 41.5 & 97.0 & 16.0\% & 61 \\
Exercise 1 code quality & 83 & 84.63 & 6.89 & 88.0 & 8.0 & 62 & 95 & 26.5\% & 12 \\
Exercise 2 code quality & 162 & 90.65 & 4.85 & 92.0 & 6.5 & 68 & 98 & 74.7\% & 14 \\
Exercise 3 code quality & 216 & 81.23 & 7.53 & 82.0 & 7.0 & 42 & 95 & 12.0\% & 17 \\
Exercise 1 automated tests & 83 & 98.93 & 1.87 & 100.0 & 1.5 & 89 & 100 & 98.8\% & 7 \\
Exercise 2 automated tests & 162 & 99.12 & 2.31 & 100.0 & 2.0 & 76 & 100 & 99.4\% & 7 \\
Grade point average & 214 & 82.87 & 6.12 & 83.0 & 8.47 & 64.86 & 98.30 & --- & --- \\
\bottomrule
\end{tabular}
\end{adjustbox}
\end{table}

Table~\ref{tab:distributions} describes the four measures on the usable records. The oral score is widely spread on all three exercises, taking 31, 33 and 26 distinct values, with standard deviations of 12.55 on exercise 1, 12.57 on exercise 2 and 8.79 on exercise 3. It therefore separates students rather than placing them in a single passing band. The code quality scores are narrower on every exercise, with standard deviations between 4.9 and 7.5 and between 12 and 17 distinct values. They are produced by an agent that reviews the submitted code and never sees the examination, and code that compiles and meets the brief leaves less room to differentiate students than their account of it does. The automated test scores of the same submissions are even narrower, with standard deviations of 1.87 and 2.31, seven distinct values on each exercise, and 98.8\% and 99.4\% of students at or above 90.

The oral score on exercise 3 is compressed relative to the other two exercises. Its median is 88 against 80 and 81, and 43.5\% of students score at or above 90, against 12.0\% and 7.4\% on exercises 1 and 2. Its scores still range from 48 to 97, but bunch more tightly at the top. Three explanations are available, and we do not choose between them. First, exercise 3 was the only exercise on which both groups were assigned the AI examination, and for almost all who sat it, it was their second, so almost all came to it with prior exposure to the format. Second, the topic pool for that exercise was narrower: its examinations drew on 20 distinct topics, against 43 and 30 on the other two. Third, exercise 3 was the one exercise without an automatic grader, so the incentives around the submission itself differed. The compression did not remove the relationship that S2 tests: exercise 3 produced the strongest single-exercise correlation with academic standing (Section~\ref{sec:standing}).

\subsection{The oral score and academic standing}\label{sec:standing}

On exercise 3, on which S1 to S3 are evaluated, the oral score and the code quality score of the same submission are uncorrelated, at −0.073 with interval [−0.205, +0.063] at n = 216, an upper bound well below the 0.30 that S1 requires. The oral score correlates +0.419 with degree grade point average, with interval [+0.294, +0.523] at n = 214, clearing the lower bound of 0.10 that S2 requires. The difference between the two correlations is +0.489 with interval [+0.312, +0.662] at n = 214, which excludes zero in the positive direction as S3 requires.\footnote{S3 is computed on the 214 students for whom both correlations are available. On that sample the oral score against the code quality score is −0.070, which is why S3 differs slightly from the difference between the two figures above.}

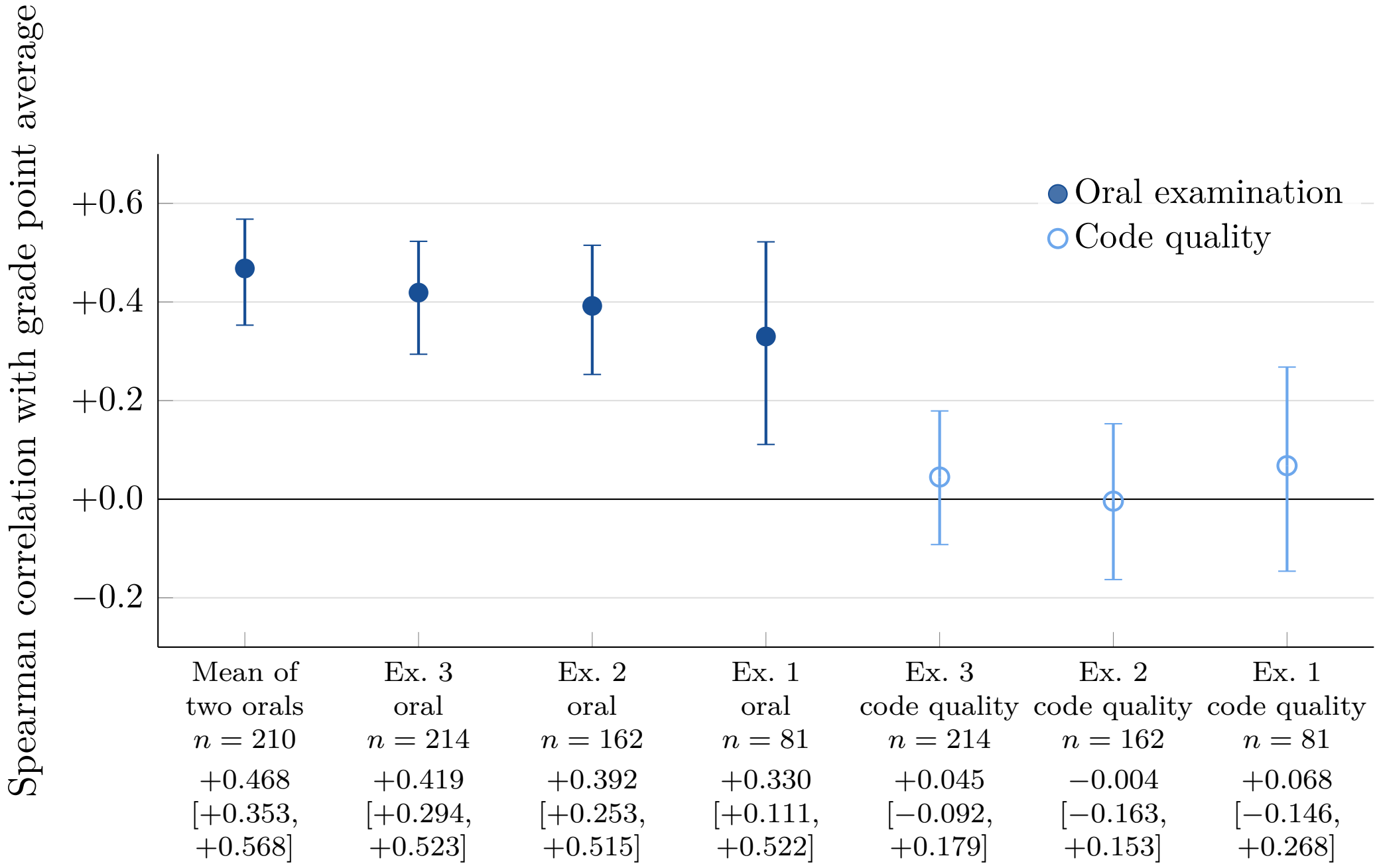
\begin{figure}[t]
\centering
\begin{tikzpicture}
\begin{axis}[
  width=\linewidth, height=6.5cm,
  ymin=-0.3, ymax=0.7,
  ytick={-0.2,0,0.2,0.4,0.6},
  ylabel={Spearman correlation with grade point average},
  xmin=0.5, xmax=7.5,
  xtick={1,...,7},
  xticklabels={
    {Mean of\\two orals\\$n=210$\\[2pt]$+0.468$\\{[}$+0.353$,\\$+0.568${]}},
    {Ex.\ 3\\oral\\$n=214$\\[2pt]$+0.419$\\{[}$+0.294$,\\$+0.523${]}},
    {Ex.\ 2\\oral\\$n=162$\\[2pt]$+0.392$\\{[}$+0.253$,\\$+0.515${]}},
    {Ex.\ 1\\oral\\$n=81$\\[2pt]$+0.330$\\{[}$+0.111$,\\$+0.522${]}},
    {Ex.\ 3\\code quality\\$n=214$\\[2pt]$+0.045$\\{[}$-0.092$,\\$+0.179${]}},
    {Ex.\ 2\\code quality\\$n=162$\\[2pt]$-0.004$\\{[}$-0.163$,\\$+0.153${]}},
    {Ex.\ 1\\code quality\\$n=81$\\[2pt]$+0.068$\\{[}$-0.146$,\\$+0.268${]}}},
  xticklabel style={align=center,font=\scriptsize},
  yticklabel style={font=\footnotesize,/pgf/number format/.cd,fixed,fixed zerofill,precision=1,showpos},
  ylabel style={font=\small},
  ymajorgrids, grid style={gray!25},
  axis x line*=bottom, axis y line*=left,
  legend style={at={(0.99,0.98)},anchor=north east,font=\footnotesize,draw=none,fill=white,fill opacity=0.8,text opacity=1},
  legend cell align=left,
]
\addplot[black, thin, forget plot] coordinates {(0.5,0) (7.5,0)};
\addplot[only marks, mark=*, mark size=2.6pt, color=atcdark,
  error bars/.cd, y dir=both, y explicit, error bar style={line width=0.8pt}, error mark options={rotate=90,mark size=2.5pt}]
  coordinates {
  (1,0.468) -= (0,0.115) += (0,0.100)
  (2,0.419) -= (0,0.125) += (0,0.104)
  (3,0.392) -= (0,0.139) += (0,0.123)
  (4,0.330) -= (0,0.219) += (0,0.192)
  };
\addlegendentry{Oral examination}
\addplot[only marks, mark=o, mark size=2.6pt, color=atclight, thick,
  error bars/.cd, y dir=both, y explicit, error bar style={line width=0.8pt}, error mark options={rotate=90,mark size=2.5pt}]
  coordinates {
  (5,0.045) -= (0,0.137) += (0,0.134)
  (6,-0.004) -= (0,0.159) += (0,0.157)
  (7,0.068) -= (0,0.214) += (0,0.200)
  };
\addlegendentry{Code quality}
\end{axis}
\end{tikzpicture}
\caption{Spearman correlation of each score with degree grade point average, with 95\% bootstrap intervals. Only the exercise 3 oral coefficient is a test of a criterion (S2), and the others are exploratory. The oral score correlates with grade point average in every sample in which it was measured, and the code quality score of the same submissions does not.}
\label{fig:gpa}
\end{figure}

Figure~\ref{fig:gpa} shows the correlation of each score with grade point average on every exercise, and that of the mean of two orals. The first result is that the oral score tracks academic standing, and not only on the exercise on which S2 was evaluated. On exercise 1 the estimate is +0.330 [+0.111, +0.522] at n = 81, and on exercise 2 it is +0.392 [+0.253, +0.515] at n = 162. The two samples are almost disjoint, sharing only two students, so the relationship holds in two separate groups rather than resting on one. The mean of two orals correlates +0.468, with interval [+0.353, +0.568] at n = 210. For the one student with two earlier examinations, it uses the oral score from that student's own told-AI exercise. Averaging two sittings draws on more evidence (Section~\ref{sec:reliability}), and the mean is the quantity that the benchmark in Section~\ref{sec:benchmark} compares against a course grade.

Degree grade point average is not on a common scale across degree tracks: the track means in this cohort range from 78.1 to 87.9. On the pure computer science track alone, the oral score correlates with grade point average at +0.433 [+0.268, +0.574] on exercise 3 (n = 122) and +0.490 [+0.335, +0.620] for the mean of two orals (n = 119), higher than the +0.419 and +0.468 of the full sample. This is the largest track, since the other named tracks have between 23 and 33 students each, too few for separate coefficients. It also places every student on a single grade scale, and its degree courses are closest to programming. On a track such as computational biology, grade point average also reflects courses in other subjects, in which a student may do better or worse than in programming.

The second result is that the code quality score does not track academic standing. Its coefficients are +0.068, −0.004 and +0.045 on exercises 1 to 3, and every interval spans zero (Figure~\ref{fig:gpa}). Taken with S1, the two scores behave differently on the same submission: they rank students differently, and only one is related to academic standing.

\subsection{The automated test scores: a ceiling, and an incentive null}\label{sec:ceiling}

S4 and S5 are evaluated on exercises 1 and 2, the two exercises that had an automatic grader (Table~\ref{tab:incentive}). The premise that S4 tests, that the course's automatic tests discriminate too little between students to carry a grade (Section~\ref{sec:analysis}), came from the course's previous year, in which 62.5\% of 272 students averaged exactly 100 over four autograded exercises, 92\% averaged 90 or above, and the median was 100.

\begin{table}[t]
\centering
\small
\caption{Automated test scores on the two exercises that had an automatic grader, by incentive arm. S4 is evaluated on the told-autograde rows. S5 compares the two rows of each exercise.}
\label{tab:incentive}
\begin{adjustbox}{max width=\textwidth}
\begin{tabular}{lcccccc}
\toprule
 & n & median & IQR & mean & SD & at 100 \\
\midrule
Exercise 1, told-AI & 84 & 100 & 2 & 98.13 & 7.43 & 51.2\% \\
Exercise 1, told-autograde & 158 & 100 & 2 & 96.82 & 13.28 & 51.9\% \\
Exercise 2, told-AI & 163 & 100 & 2 & 99.11 & 2.30 & 71.2\% \\
Exercise 2, told-autograde & 84 & 100 & 2 & 99.12 & 1.74 & 73.8\% \\
\bottomrule
\end{tabular}
\end{adjustbox}
\end{table}

\begin{figure}[t]
\centering
\pgfplotsset{distplot/.style={
  width=\linewidth, height=5.4cm,
  ybar, bar width=13pt, ymin=0, ymax=105,
  symbolic x coords={a,b,c,d,e}, xtick=data,
  xticklabels={Below 80,80--89,90--94,95--99,Exactly 100},
  xticklabel style={font=\scriptsize,align=center,text width=1cm},
  yticklabel style={font=\footnotesize},
  title style={font=\footnotesize,align=center},
  nodes near coords, nodes near coords style={font=\scriptsize},
  axis x line*=bottom, axis y line*=left, enlarge x limits=0.15}}
\begin{minipage}[t]{0.48\linewidth}
\centering
\begin{tikzpicture}
\begin{axis}[distplot, ylabel={Students}, ylabel style={font=\small},
  title={Automated tests, exercise 2,\\told-autograde ($n=84$)}]
\addplot[fill=atclight, draw=atclight] coordinates {(a,0) (b,0) (c,2) (d,20) (e,62)};
\end{axis}
\end{tikzpicture}
\end{minipage}\hfill
\begin{minipage}[t]{0.48\linewidth}
\centering
\begin{tikzpicture}
\begin{axis}[distplot, title={Oral examination, exercise 3\\($n=216$)}]
\addplot[fill=atcdark, draw=atcdark] coordinates {(a,30) (b,92) (c,70) (d,24) (e,0)};
\end{axis}
\end{tikzpicture}
\end{minipage}
\caption{Distribution of scores in two instruments. Left: the automated test scores of the students told their exercise 2 submission would be graded automatically take five distinct values, and 62 of the 84 sit at exactly 100. Right: oral examination scores on exercise 3 take 26 distinct values, from 48 to 97.}
\label{fig:distributions}
\end{figure}
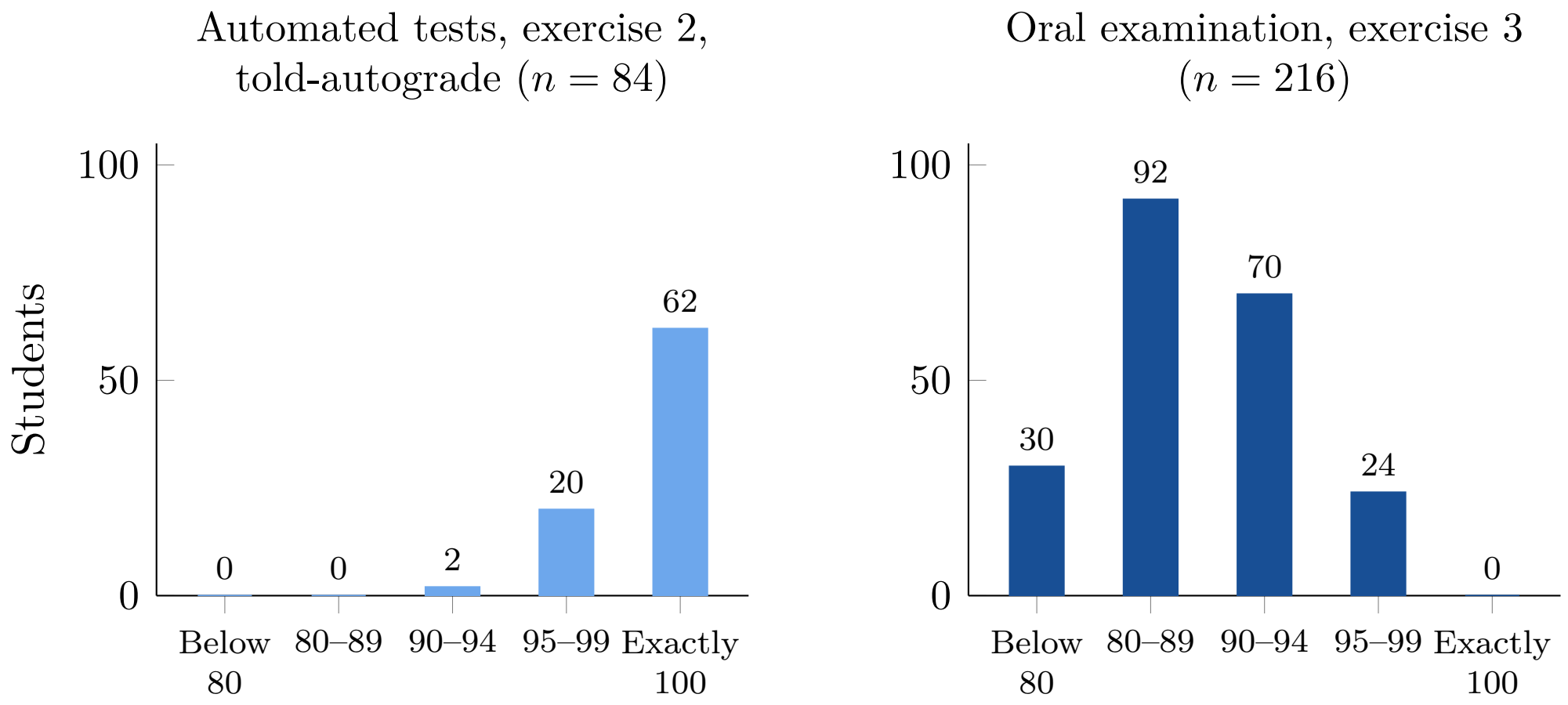

On the told-autograde arms both exercises meet S4, with a median of 100 and an interquartile width of 2. Exercise 2 is the starker case: its 84 students receive five distinct scores, the lowest of which is 93, the standard deviation is 1.74, and 73.8\% of them sit at exactly 100. Exercise 1 has a larger standard deviation, 13.28, but it comes entirely from a low tail reaching 9, across 13 distinct values. Its central half still lies between 98 and 100, and 51.9\% of its students sit at exactly 100. Because these scores count exactly matching outputs (Section~\ref{sec:exercises}), a low score can mean correct logic printed in a non-conforming format rather than a failure of understanding. Figure~\ref{fig:distributions} sets the exercise 2 distribution beside the oral score distribution on exercise 3.

\label{sec:incentive}Table~\ref{tab:incentive} also carries the incentive contrast that S5 tests. Within each exercise both arms were scored by the same harness but had been told different things about how their submissions would be graded. S5 is not met. All four medians sit at exactly 100, so the median difference is 0.0 on both exercises. Its bootstrap interval is [−1.0, +1.0] on exercise 1 and degenerate on exercise 2, where all 10,000 replicates returned exactly zero because resampling never moves either median off 100. Both are percentile intervals, since the BCa adjustment cannot be computed when every leave-one-out median is 100. Two rank-based checks also find no difference. Cliff's delta, which measures how often a score from one arm exceeds a score from the other, on a scale from −1 to +1, is −0.0092 on exercise 1 and +0.0148 on exercise 2, with intervals spanning zero, and the Mann--Whitney test gives p = 0.899 and p = 0.810.

Three features limit what this null can show. First, the harnesses are saturated in both arms, at a median of 100 and an interquartile width of 2 in all four cells, so the S4 ceiling holds on the told-AI side as well. A difference in incentive cannot be detected in a score that half or more of every arm has already carried to its maximum. Second, the two exercises used different harnesses (Section~\ref{sec:exercises}), so these are two separate comparisons on two instruments rather than replications of one. Third, the told-AI students still had a stake in their code, since code quality carried 25\% of their grade, so the comparison sets two incentives against each other rather than an incentive against none. The null therefore does not show that incentives leave no mark on how students write code.

The ceiling has a second consequence, which appears when the automated test score is set against academic standing (Table~\ref{tab:autogpa}). On exercise 2 the two are related, at +0.141 for all students and +0.261 in the told-autograde arm, both with intervals that exclude zero. On that exercise, however, nearly three quarters of each arm sit at exactly 100, and the told-autograde arm takes only five distinct values. An instrument with that little spread can barely rank students, so a coefficient between it and academic standing is not evidence of a stable relationship. On exercise 1 every interval spans zero. Across the six cells, the automated test scores therefore show no consistent relationship with academic standing.

\begin{table}[t]
\centering
\small
\caption{Spearman correlation between the automated test score and degree grade point average, by exercise and arm.}
\label{tab:autogpa}
\begin{adjustbox}{max width=\textwidth}
\begin{tabular}{lccc}
\toprule
 & n & Spearman & 95\% CI \\
\midrule
Exercise 1, all students & 240 & −0.045 & [−0.171, +0.080] \\
Exercise 1, told-autograde & 158 & −0.060 & [−0.214, +0.099] \\
Exercise 1, told-AI & 82 & −0.014 & [−0.235, +0.203] \\
Exercise 2, all students & 245 & +0.141 & [+0.018, +0.256] \\
Exercise 2, told-autograde & 82 & +0.261 & [+0.018, +0.454] \\
Exercise 2, told-AI & 163 & +0.081 & [−0.061, +0.223] \\
\bottomrule
\end{tabular}
\end{adjustbox}
\end{table}

The students who sat the oral examination on exercises 1 and 2 also allow the three scores to be compared on identical submissions (Table~\ref{tab:samesub}). The oral score shows no detectable correlation with the code quality score on either exercise, at −0.014 and −0.106 with intervals spanning zero, which is consistent with S1 on exercise 3. Nor does it correlate detectably with the automated test score, at −0.084 and +0.047. The two scores that read the submitted code are related on exercise 1, at +0.573 with interval [+0.396, +0.702], and not on exercise 2, at +0.017 with interval [−0.146, +0.186], where most students sit at 100 and there is little for a rank coefficient to order (\ref{app:stats}). On exercise 1, then, the two code-based scores move together and the oral score moves with neither. Exercise 2 does not repeat that, so it is one observation rather than a pattern. Because the automated test scores are concentrated at the top of their range, range restriction attenuates every coefficient involving them in Table~\ref{tab:samesub}, and these coefficients are likely to understate whatever association exists in a population with more variance. We apply no correction, since one would require assumptions about the unrestricted distribution that these data cannot support.

\begin{table}[t]
\centering
\small
\caption{Spearman correlations among the three scores on the same submissions, for the students who sat the oral examination on exercises 1 and 2. Exploratory: computed after the success criteria were evaluated. Intervals are 95\% bootstrap intervals.}
\label{tab:samesub}
\begin{adjustbox}{max width=\textwidth}
\begin{tabular}{lcccc}
\toprule
 & \multicolumn{2}{c}{Exercise 1 ($n = 83$)} & \multicolumn{2}{c}{Exercise 2 ($n = 162$)} \\
\cmidrule(lr){2-3}\cmidrule(lr){4-5}
 & Spearman & 95\% CI & Spearman & 95\% CI \\
\midrule
Oral and code quality & −0.014 & [−0.230, +0.194] & −0.106 & [−0.255, +0.053] \\
Oral and automated tests & −0.084 & [−0.297, +0.137] & +0.047 & [−0.115, +0.208] \\
Code quality and automated tests & +0.573 & [+0.396, +0.702] & +0.017 & [−0.146, +0.186] \\
\bottomrule
\end{tabular}
\end{adjustbox}
\end{table}

\subsection{Robustness checks}\label{sec:robustness}

Six checks bear on the results above, and \ref{app:robustness} reports each in full.
\begin{enumerate}
\item \emph{Answer length.} The oral score rises with the total number of words the student typed in the sitting, at +0.437, +0.385 and +0.373 on exercises 1 to 3, while the code quality score does not. Controlling for that length lowers the oral score's correlation with grade point average on exercise 3 from +0.415 to +0.349 [+0.216, +0.464], both on the same 213 students, on exercise 2 from +0.392 to +0.343, and on exercise 1 from +0.324 to +0.222. Length therefore accounts for part of that relation but not most of it.
\item \emph{Timeouts and group.} Excluding the 19 exercise 3 sittings in which time ran out leaves S1 to S3 on the same side of their thresholds, and on exercise 3 the grade distributions of groups S and L do not differ detectably.
\item \emph{Consistency of the code quality score.} A student's code quality score on exercise 3 correlates +0.388 with their code quality score on their earlier exercise, on the same 212 pairs of sittings as the oral scores in Section~\ref{sec:reliability}. The code quality score's lack of relation to grade point average is therefore unlikely to reflect instability alone.
\item \emph{Style assessment.} The Grader receives the Code Reviewer's assessment of how machine-generated the code's style looks (Section~\ref{sec:authorship}). The code quality score rises with that assessment and the oral score falls slightly with it, which may reflect the assessment's influence on the Grader as well as anything about the students. Holding the assessment constant leaves the oral and code quality scores uncorrelated on all three exercises, so S1 does not depend on it.
\item \emph{Order of sitting.} Later sitters did not score higher in aggregate (on exercise 2 they scored slightly lower), so no advantage from the recurrence of question topics within an exercise is visible.
\item \emph{Corrected code quality scores.} On the exercise 3 code quality scores corrected for the recurring review error described in Section~\ref{sec:operation}, S1 is +0.002 and S3 +0.412, so both keep their verdicts.
\end{enumerate}

\subsection{Reliability of a single examination}\label{sec:reliability}

\paragraph{Grading of a fixed transcript}
To measure how much of a score is the grading model's run-to-run variation, the Grader and, separately, the Code Reviewer were run again five times each on a stratified sample of 50 usable sittings: 15 from exercise 1, 15 from exercise 2 and 20 from exercise 3, drawn evenly from the five quintiles of the oral score within each exercise. Each repeat used the same instructions, model, temperature and inputs that had produced the stored grade, and the Grader received the stored style assessment unchanged.

Run to run, the Grader's oral scores agree closely. Across the 50 sittings the intraclass correlation of the five repeats (one-way, single run) is 0.978, with a bootstrap interval over sittings of [0.963, 0.990]. The pooled within-sitting standard deviation is 1.7 points, and the median range of a sitting's five scores is 1 point. All five repeats fall within 5 points of one another in 44 of the 50 sittings. Of the 250 repeats, 241 fall within 5 points of the stored score and 149 exactly on it, with a mean difference of 0.0 points. In the six wider sittings, with ranges of 6 to 12 points, a half-point difference in re-scoring a single answer moves the average across a band of the score table. Three of them sit near an oral score of 60. Exercise 3 is the most stable, with a within-sitting standard deviation of 1.0 point, no range above 6 points, and every repeat within 5 points of the stored score. The Code Reviewer is almost deterministic. Across 49 of the same sittings its intraclass correlation is 0.999 and its within-sitting standard deviation 0.3 points. No sitting's range exceeds 3 points, and every repeat falls within 5 points of the stored score. When both agents vary together, on those 49 sittings, the combined grade has a within-sitting standard deviation of 1.3 points.

Run-to-run variation in grading is therefore small compared with the spread of the oral score, whose standard deviation is 8.8 to 12.6 points (Section~\ref{sec:distributions}), and it cannot account for the correlations reported in Section~\ref{sec:standing}. It measures only the grading of a fixed transcript. A second examination of the same student would bring different questions, drawn and phrased afresh, and different answers, so its outcome could vary more than the grading does. The correlation between two sittings, reported next, bears on that.

\paragraph{Two examinations of the same student}
A student's two oral scores correlate +0.313 with each other (n = 212, 95\% interval [+0.182, +0.432]). The two sittings examined different exercises, a median of three weeks apart. This figure therefore mixes measurement error with real differences in how well a student understood two different submissions, and the data do not separate the two. For comparison, two five-minute oral examinations given by staff in the course of 628 students described in Section~\ref{sec:related} correlated 0.34 within students \citep{ellis2026scale}. Controlling for grade point average lowers the agreement to +0.193 [+0.059, +0.317] (n = 210, against +0.317 without the control on the same students), so the agreement is not only a reflection of academic standing. The agreement between a student's two code quality scores is barely changed by the same control, at +0.396 without it and +0.401 with it (n = 210). How much of the figure is error cannot be read from it, so the reliability of a single 16-minute sitting remains unestimated. This is why Section~\ref{sec:standing} reports the mean of two alongside the single-exercise figures: it rests on more evidence.

\subsection{Benchmark against course grades}\label{sec:benchmark}

\begin{figure}[t]
\centering
\begin{tikzpicture}
\begin{axis}[
  width=0.66\linewidth, height=5.4cm,
  xmin=0, xmax=1.0,
  xtick={0,0.2,...,0.6},
  clip=false,
  xticklabel style={font=\footnotesize,/pgf/number format/.cd,fixed,fixed zerofill,precision=1,showpos},
  xlabel={Spearman correlation with the remaining grade point average}, xlabel style={font=\small},
  ymin=0.5, ymax=5.5, ytick={1,...,5}, y dir=reverse,
  yticklabels={
    {Object-oriented programming\\course grade (one semester)},
    {Mean of two oral examinations\\(32 minutes)},
    {Introduction to computer science\\course grade (one semester)},
    {Exercise 3 oral alone\\(16 minutes)},
    {Computer architecture\\course grade (one semester)}},
  yticklabel style={align=right,font=\scriptsize},
  xmajorgrids, grid style={gray!25},
  axis x line*=bottom, axis y line*=left,
  legend style={at={(0.5,1.03)},anchor=south,legend columns=2,font=\footnotesize,draw=none,/tikz/every even column/.append style={column sep=0.5cm}},
  legend cell align=left,
]
\addplot[only marks, mark=*, mark size=2.6pt, color=atcdark,
  error bars/.cd, x dir=both, x explicit, error bar style={line width=0.8pt}, error mark options={mark size=2.5pt}]
  coordinates {(0.453,2) -= (0.161,0) += (0.132,0)
               (0.386,4) -= (0.176,0) += (0.143,0)};
\addlegendentry{Oral examination}
\addplot[only marks, mark=square, mark size=2.4pt, color=atclight, thick,
  error bars/.cd, x dir=both, x explicit, error bar style={line width=0.8pt}, error mark options={mark size=2.5pt}]
  coordinates {(0.529,1) -= (0.141,0) += (0.117,0)
               (0.453,3) -= (0.157,0) += (0.133,0)
               (0.331,5) -= (0.159,0) += (0.149,0)};
\addlegendentry{Course grade}
\node[anchor=west,font=\scriptsize] at (axis cs:0.70,1) {$+0.529$ [$+0.388$, $+0.646$]};
\node[anchor=west,font=\scriptsize] at (axis cs:0.70,2) {$+0.453$ [$+0.292$, $+0.585$]};
\node[anchor=west,font=\scriptsize] at (axis cs:0.70,3) {$+0.453$ [$+0.296$, $+0.586$]};
\node[anchor=west,font=\scriptsize] at (axis cs:0.70,4) {$+0.386$ [$+0.210$, $+0.529$]};
\node[anchor=west,font=\scriptsize] at (axis cs:0.70,5) {$+0.331$ [$+0.172$, $+0.480$]};
\end{axis}
\end{tikzpicture}
\caption{Spearman correlation with degree grade point average on a common sample of 121 students, with 95\% bootstrap intervals. Each course grade is set against the grade point average without that course, and the oral scores against the full grade point average, which does not include this course. Each interval overlaps every other, so the figure places the five measures in a common range without ordering them or showing them to be equivalent.}
\label{fig:benchmark}
\end{figure}
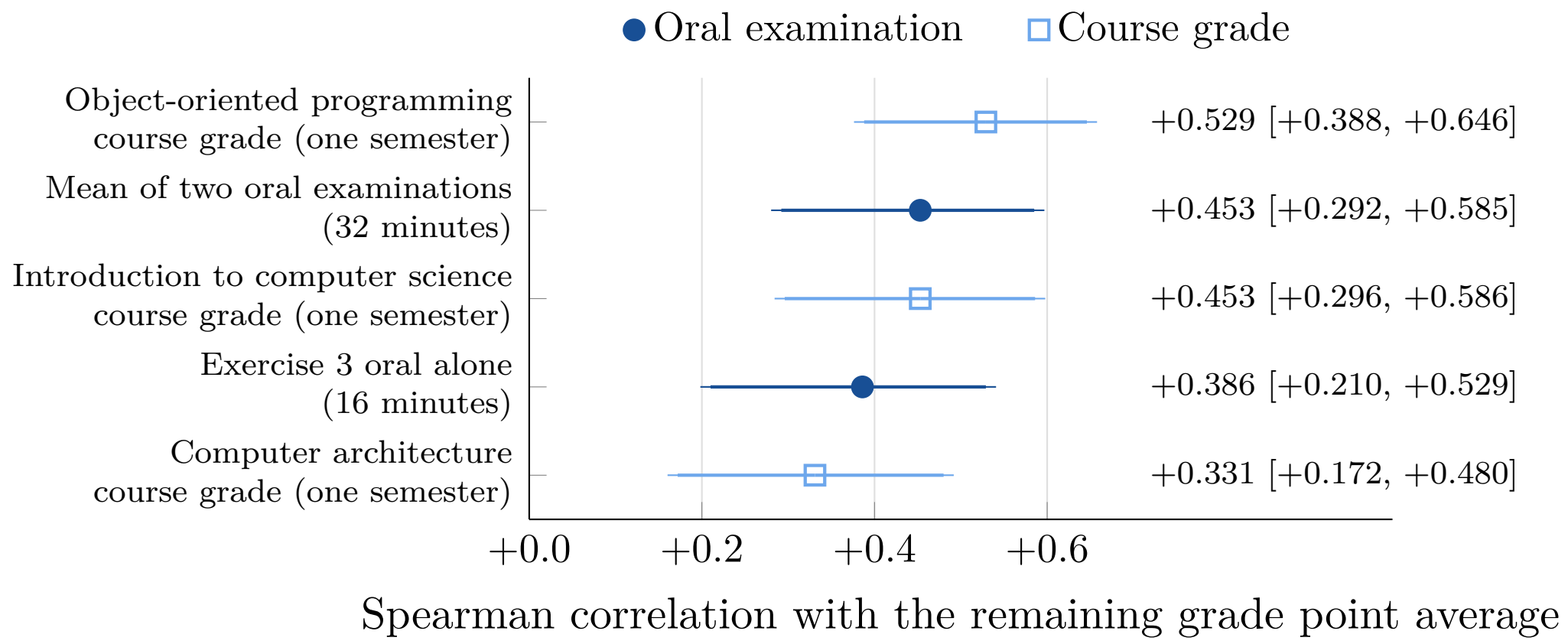

A student's two oral examinations together occupy 32 minutes of standard examination time. The natural comparison is a full-semester course grade, the familiar indicator of academic standing in this setting. We use three courses that, like this one, are built on programming assignments: object-oriented programming, introduction to computer science and computer architecture. The comparison in Figure~\ref{fig:benchmark} is exploratory. It runs on the 121 students who have a grade point average, both oral scores and all three course grades, so that every measure is estimated on the same students. By the time they take this course, students have completed 12 to 15 courses in the degree. Because each of the three courses is one of them, its grade is set against the student's grade point average with that grade removed. No student on the computational biology or software engineering track has all three course grades, so the sample is drawn from the other tracks.

On this sample, the coefficient for two 16-minute oral examinations falls within the range of those for single-semester course grades. Each interval overlaps every other, so the figure does not show that the oral examination ranks above or below any particular course, nor that it is equivalent to one.

Two asymmetries in this comparison favor the course grades. First, the course grades share their method and their institutional grading culture with the grade point average they are set against, which inflates their coefficients, whereas the oral examination shares neither. Second, the oral figure rests on two short sittings that correlate only +0.313 with each other (Section~\ref{sec:reliability}), whereas a semester course grade aggregates many assessments. Measurement error lowers a correlation, so it pulls the oral coefficient down more than it pulls down those of the course grades. Even so, 32 minutes of examination placed the oral score within the range of course grades that each rest on a semester of assignments, a quiz and a final examination of two to three hours.

\subsection{How students evaluated the two modes}\label{sec:reception}

The surveys (Section~\ref{sec:surveys}) carried ten hypotheses about how students would evaluate the two modes, numbered H1 to H10. \ref{app:reception} states each hypothesis with its decision rule and gives the response rates and the full results. Three hypotheses were designated primary and are reported with 98.33\% intervals, the 95\% level adjusted for three tests (Bonferroni). The other seven are exploratory and are reported with 95\% intervals. They concern attitudes rather than learning, grade quality or how well either mode measures understanding. Because only the AI examination had a post-examination form, every comparison between the two modes sets expectations against expectations.

All three primary hypotheses are supported. First, before experiencing either mode, students expected the autograder's grade to reflect their understanding and mastery better than the AI examination's (H1, the naive expectation gap). The comparison rests on two items worded identically on both pre-examination forms, combined into one composite score, and gives a Cliff's delta of +0.305 with a 98.33\% interval of [+0.093, +0.490], on 103 respondents to the autograder form and 74 to the AI form at exercise 1. Second, students rated the AI examination more positively after the exercise 3 examination than before it (H3, the exercise-three experience gain), with a matched-pairs rank-biserial correlation of +0.489 [+0.214, +0.699] on 107 complete pairs. Third, support for keeping an AI-based component in grading rose (H4, the retention-support shift). The net switch, the share of students who moved toward keeping such a component minus the share who moved away, is +0.159 [+0.034, +0.273] on 88 pairs. One exploratory result runs the other way. Students meeting the AI examination first, at exercise 1, rated it lower afterwards than they had expected beforehand (H7), at −0.349 with a 95\% interval of [−0.603, −0.064] on 63 pairs. The exercise 3 gain does not repeat for either group at its first encounter with the AI examination, which was exercise 1 for group S and exercise 2 for group L. \ref{app:reception} also sets out what qualifies each supported result: H1 is confounded with lecture group, H3 and H4 were answered immediately after a grade from the system, and the students who answered both forms on exercise 3 had higher oral scores.

\subsection{What the deployment cost}\label{sec:cost}

Examining the 546 sittings cost an average of at most \$1.37 per sitting in model spend through the Claude API, at a listed \$5 and \$25 per million input and output tokens for Claude Opus 4.6, and \$198.38 in cloud infrastructure on Google Cloud Platform for the whole deployment, with no trial allowance or institutional grant (Table~\ref{tab:cost}). Running the deployment took about 40 hours of a teaching assistant's time, covering setup, GitHub Classroom intake, roster reconciliation, and student support and troubleshooting between exercises. Building the system with Claude Code took a further 80 hours or so, including quality assurance, testing and debugging, and \$153.34 of model spend, once. Both hour figures are the teaching assistant's estimates from recollection. Verifying the topic pool took the teaching assistant and the co-instructor two to three hours per exercise, and the course instructor's sampled review of the recordings took about 6 hours. Authoring the exercises is not counted, since it belongs to setting an assignment, and no hours went to attending the examination room, which was unattended.

\begin{table}[t]
\centering
\small
\caption{What the deployment cost. Measured and estimated quantities are kept on separate rows and are not combined.}
\label{tab:cost}
\begin{adjustbox}{max width=\textwidth}
\begin{tabular}{>{\raggedright\arraybackslash}p{4.3cm}>{\raggedright\arraybackslash}p{3.7cm}>{\raggedright\arraybackslash}p{3.6cm}l}
\toprule
Quantity & Amount & Basis & Status \\
\midrule
Model spend, examination days & \$747.19, an average of at most \$1.37 per sitting & Billing export, 546 sittings & Measured \\
Model spend, development and debugging & \$153.34, one-off & Billing export & Measured \\
Cloud infrastructure & \$198.38, whole deployment & Billing export & Measured \\
Building the system & About 80 hours, one-off & Teaching assistant's recollection & Estimated \\
Running the deployment & About 40 hours & Teaching assistant's recollection & Estimated \\
\bottomrule
\end{tabular}
\end{adjustbox}
\end{table}

For scale, a human examiner giving the same 546 sittings at the same 16 minutes would spend about 146 hours in examination alone, before reading any code, grading or writing feedback. In the course of 628 students described in Section~\ref{sec:related}, two five-minute oral examinations per student took 21 staff an estimated 180 hours \citep{ellis2026scale}. This comparison is hypothetical: no human examination of this cohort was run, and it concerns cost rather than the quality of the judgment (Section~\ref{sec:limitations}). Model prices for a given level of performance have fallen by between 9 and 900 times per year, depending on the task \citep{cottier2025inference}, so the order of magnitude of these figures, and the human hours, matter more than the exact amounts.

\section{Discussion}\label{sec:lessons}

This study asked two questions of a system that examines each student on their own code: whether it can run on a full course, and whether its scores relate to other measures as its design predicts. In this deployment the answer to both is yes. The system examined 247 students in 546 sittings with no examiner in the room, and the oral score it produced was unrelated to the quality of the code it examined while tracking how students stood in the rest of their degree. The subsections below take up what the oral score shows, what the course's automatic tests could not show, what running the system required, how many sittings a grade should rest on, how students received the examination, and what assessing understanding means when AI use is permitted.

\subsection{What the oral score shows}\label{sec:disc-oral}

The central result of the previous section is a contrast between two scores that the same system computes from the same submission. The code quality score and the oral score rank students differently, and only the oral score is related to how students stand elsewhere in the degree, in every sample in which it was measured (Section~\ref{sec:standing}). The scores computed from the submitted code alone, by a reviewing agent or by an automatic test harness, showed no consistent relation to academic standing, while the score from asking students to account for that code did. One reading, which these data do not test, is that where AI assistance is permitted, as it was here, the quality of the code says little about the student who handed it in, while the student's account of it says more. No human examiner scored these students, so neither instrument was checked against a standard, and what the oral score tracks remains to be established.

One may wonder whether the lack of correlation between the two scores, which we call their decorrelation, is an artifact of the architecture. The two scores come from the same system, and their low correlation might reflect how the system splits its own judgment across two axes rather than anything about students. Two features of the design argue against this. First, the two scores are not two slices of one judgment. The Code Reviewer sees the source and never the transcript, the Grader sees the transcript and never receives the code quality score, and the Code Reviewer runs first and receives nothing from the Grader, so neither can adjust its score toward or away from the other's. Three code-derived routes do reach the Grader (Section~\ref{sec:grading}): the lines the examiner quoted in its questions, the per-turn scores recorded by an examiner that had read the code, and the Code Reviewer's style assessment, which the Grader receives for the authorship decision. The first two would most plausibly pull the oral score toward the code quality score. The two scores relate to the third in opposite directions (Section~\ref{sec:robustness}), so it could push their correlation down. With that signal held constant, however, the correlation stays near zero on all three exercises (\ref{app:robustness}). Second, the decorrelation does not stand alone. The oral score converges with degree grade point average, a measure produced entirely outside the system, over courses it never saw, and a split of the system's own judgment would not by itself produce that convergence. A grader that rewarded features of the writing that also go with academic standing, such as length or fluency, could. In our analysis, answer length was measured and accounts for only a small part of the convergence (Section~\ref{sec:robustness}). Taken together, these checks leave the decorrelation standing.

\subsection{What automated tests can carry}\label{sec:disc-tests}

The automated test scores show no consistent relation to academic standing because their distributions have so little spread that the instrument barely separates students. The weak and inconsistent correlations there are therefore a property of its variance rather than evidence of independence (Section~\ref{sec:ceiling}). The same saturation is the main reason S5 could not detect a difference. S5 asked whether students who expected automatic grading aimed their code at the tests, and so scored higher on them than students who expected the AI examination. Half or more of every arm, however, reached the maximum, and a score at its maximum cannot register such a difference. The failed criterion therefore says more about the instrument than about the students. The ceiling under automatic grading was anticipated, since S4 was fixed in advance and met. That it held in the told-AI arm as well was not. The saturation describes these test suites on these exercises rather than automated testing in general, and tests built to separate students might well do so.

For a course that relies on automatic tests, the transferable point is what such a score can carry once it saturates: it can show that code runs, but near the ceiling it ranks students only coarsely. The oral score separates students whom the tests do not. On exercise 2, where 99.4\% of the submissions scored 90 or above on the tests, the oral scores of the same students ranged from 15 to 93 (Table~\ref{tab:distributions}). This is in line with the observation, made before generative AI, that about a third of introductory students whose programs passed the automated tests struggled to explain their own code \citep{lehtinen2021struggle}. Where a model can produce code that passes the tests, a passing score certifies the code rather than the student who submitted it.

\subsection{Running the examination at class scale}\label{sec:disc-scale}

What running the examination required is the practical side of the feasibility question. The system examined 247 students in 546 sittings on a single workstation, in two examination languages, and the examiner needed its fallback model on only 3 of 2,192 calls (Section~\ref{sec:operation}). Most students (212 of the 247) had a graded examination on exactly two exercises, and without the gradual rollout each would have been examined on all three (Section~\ref{sec:crossover}). Model spend averaged at most \$1.37 per sitting. More important than the amount is how the human work scales. Reviewing the topic pool took two to three hours per exercise, whatever the number of students, and running the deployment took a teaching assistant about 40 hours over the whole semester (Section~\ref{sec:cost}). The examining itself, which in an oral examination grows with every student, was done by the system. A human examiner giving the same sittings would have spent about 146 hours in examination alone.

Autonomy here has a specific meaning. No member of staff was present in the examination room at any point, and no human reviewed or adjusted a grade before it was recorded. The human check came after the grade instead. Every student could appeal to the teaching assistants, who decided by reading the transcript and the code. Over the 461 cells with a graded sitting, 23 appeals were filed, 13 of them arising from one recurring error, which was corrected for every affected student, including those who had not appealed (Sections~\ref{sec:operation} and~\ref{sec:appeals}).

\subsection{One sitting or two}\label{sec:disc-sittings}

A single sitting is a short basis for a grade of record. A student's oral scores on two exercises correlate +0.313 (Section~\ref{sec:reliability}), close to the 0.34 between two oral examinations given by staff in a large course \citep{ellis2026scale}. A 16-minute examination by the system was therefore roughly as consistent across sittings as an examination by people. Here the two sittings examined different exercises, so the figure mixes measurement error with real differences in what students understood. The mean of two sittings tracks academic standing more closely than any single sitting, at +0.468 against +0.330 to +0.419, and on the pure computer science track at +0.490 against +0.433 on exercise 3 (Section~\ref{sec:standing}), as aggregating measurements would predict. For a course, the practical consequence is that a grade of record is better rested on more than one sitting, and the system makes that affordable: an additional examined exercise costs model spend and one review of its topic pool rather than examiner hours.

\subsection{How students received the examination}\label{sec:disc-reception}

Students began skeptical. Before experiencing either mode, they expected the autograder's grade to reflect their understanding better than the AI examination's (Section~\ref{sec:reception}). That expectation should be read beside the ceiling described in Section~\ref{sec:disc-tests}. Under automatic grading, half or more of every arm scored exactly 100, and with AI assistance permitted, such a grade could be reached with little effort. An examination that asks students to account for their code takes that easy grade away and adds work that automatic grading never asked for: preparing to explain the code, booking a slot and coming to the examination room. Part of the resistance to it may be resistance to that loss and that effort. In the university's end-of-course teaching evaluation, which is separate from this study, several students objected that the system had removed an easy part of a demanding course and added the time needed to prepare and attend, while others described it as the way knowledge should be assessed now that AI can write the code. After the exercise 3 examination, for almost all of them their second, students rated the AI examination more positively than they had before it, and support for keeping an AI-based component in grading rose. The first encounter did not show this gain. Students who met the AI examination first, at exercise 1, rated it lower afterwards than they had expected, and those who met it at exercise 2 showed no change. These responses were given shortly after a grade from the system, and the students who answered both forms on exercise 3 had higher oral scores than those who did not (\ref{app:reception}). With that caution, they suggest that acceptance came with a second encounter rather than the first.

\subsection{Assessing understanding when AI use is permitted}\label{sec:disc-ai}

Much of the response to generative AI in assessment concerns whether a student used it. ATC asks instead whether the student understands the code they submitted, however it was produced. Students were permitted to use AI assistants, and the grade does not depend on whether they did (Section~\ref{sec:goals}). The grade therefore rests on the student's own account of the code rather than on a judgment about how the code was written, and the design needs no detector of machine-generated code to be fair. A student could prepare by having a model explain their code to them. That preparation is itself study of the code, and the questions are designed so that recalling a prepared explanation is not enough. They are formulated during the examination from specific lines, and they include questions on why the student made a design choice and on what the code would do if an input or a condition changed, which an explanation of what the code does is unlikely to cover (Section~\ref{sec:questioning}).

Authorship is recorded but kept out of the grade. The examiner rates whether the student speaks about the code as its author, and the outcome goes to staff. The two staff reports in Section~\ref{sec:authorship} show why the axes are kept apart. Code that looked machine-generated came with an account that scored 92, and a student who could not explain a line of their own script scored 69. Merging the two axes would have penalized the first student for how the code looked, or let a suspicion of misuse decide the second student's grade. Keeping authorship advisory leaves decisions about misconduct with people, who can weigh evidence the system does not have. Together, these choices bring the oral defense that the integrity literature proposes \citep{kofinas2025} to the scale of a full course.

\section{Limitations and threats to validity}\label{sec:limitations}

The limitations below bear mainly on the second of the two claims set out at the start of Section~\ref{sec:lessons}, that the system's scores relate to other measures as its design predicts. They are ordered by how directly they bear on it, and for each we state what it affects and what it leaves standing.

The clearest limitation is the absence of human-examiner ground truth. No human examiner scored the same students on the same submissions, so we cannot state how closely the oral score tracks the grade a trained examiner would have given after a comparable defense. Where no accepted standard exists, the usual course is to test whether a score relates to other measures as theory predicts \citep{campbell1959}, and the success criteria were built for that purpose (Section~\ref{sec:analysis}). Human oral examination is itself a noisy criterion: two oral examinations given by staff in a large course correlated 0.34 within students \citep{ellis2026scale}, close to the 0.313 between two sittings here (Section~\ref{sec:reliability}). The grades also met human checks. Throughout the deployment, the course's two teaching assistants read samples of questions, answers and scores and found them sound, although without a rating scheme or a measure of agreement between them. Every student could appeal, and over the 461 cells with a graded sitting, five appeals contested the oral score. All five were upheld, so the oral score was revised in about 1\% of cells (Section~\ref{sec:operation}). Grading was also stable: re-running the Grader on a fixed transcript reproduced its scores closely, with an intraclass correlation of 0.978 (Section~\ref{sec:reliability}). Section~\ref{sec:future} describes the study that would measure agreement with human examiners.

A second limitation concerns the independence of the oral score and the code quality score. Both are model outputs, and the three agents run on the same model and share a prompting style. Separating their inputs limits contamination between transcript and code without making the pipeline independent of itself. A shared model could shape how the two scores relate to each other, but it cannot explain why the oral score relates to grade point average, a measure the system never saw (Section~\ref{sec:disc-oral}). One input crosses the separation by design: the Grader receives the Code Reviewer's style assessment for the authorship decision. Whether that input moves the oral score itself was not tested by withholding it. Whatever its influence, it does not drive the central result: holding the assessment constant leaves the oral and code quality scores uncorrelated on all three exercises (Section~\ref{sec:robustness}). A future implementation could take the authorship decision in a step that does not produce the oral score.

Degree grade point average is a coarse external criterion. It aggregates courses that differ in subject, format and difficulty, most of them unrelated to systems programming, and it is not on a common scale across programs. Coarseness of this kind lowers a correlation, so the oral score's +0.419 is a conservative estimate of its relation to academic standing. Where the scale is common, on the pure computer science track, the coefficient is higher, at +0.433 for a single sitting and +0.490 for the mean of two (Section~\ref{sec:standing}). The same criterion limits the three course grades used as benchmarks, which correlate with it at 0.33 to 0.53 (Section~\ref{sec:benchmark}), so the oral score is held to the standard that established measures meet. The analyzed sample also leaves out the weaker end of the cohort: the 21 students who submitted exercise 3 but had not sat it had a lower grade point average (median 78.7 against 83.0). A narrower range of students makes a correlation smaller, so their absence works against the result rather than for it.

The deployment is one course at one institution, with one teaching team and one submission format, and the dimensions from which topics were generated and the examiner's instructions were written for systems programming. A single deployment cannot fully separate properties of the design from properties of this cohort. Within it, however, the result repeated: the oral score tracked grade point average on each of three exercises, written in two programming languages, and in two groups of students that shared only two members on exercises 1 and 2 (Section~\ref{sec:standing}). The results also belong to one model version, Claude Opus 4.6. A newer model would be a different instrument, but the parts of the design that the results rest on, the separation of inputs and the fixed arithmetic, do not depend on the model, and the criteria fixed in advance can be applied again to a new model at the cost of one deployment. Section~\ref{sec:future} notes the further courses in which the system has since been used.

The crossover limits what any comparison between the two modes can support. Mode is tied to exercise content within each group (Section~\ref{sec:crossover}), and the two groups were formed by students' choice of lecture group rather than by randomization. They differ in size, 84 students against 163, and in track composition, with 63.8\% of group L on the pure computer science track against 42.9\% of group S (full-cohort denominators). This bears only on the secondary comparisons, of how the modes were received and of automated test scores under each incentive. It does not bear on S1 to S3, which were evaluated on exercise 3, where the whole cohort was assigned the AI examination and the two groups' oral, code quality and combined grades did not differ detectably (\ref{app:robustness}). Group is therefore treated as a descriptive stratum, and no group contrast is presented as a causal effect.

The examination room was unattended, and the recordings were reviewed by sampling rather than in full. Each student sat at a workstation locked to the examination, two cameras recorded continuously, and every sitting had a chance of being observed (Section~\ref{sec:room}), a more controlled setting than that of the closest deployed system, in which students sat the examination on their own machines wherever they chose \citep{ipeirotis2026}. The sampled review found no violation in any sitting it covered. Undetected help during a sitting would also most plausibly have raised the scores of weaker students, which would weaken the relation between the oral score and grade point average rather than produce it.

\section{Future work}\label{sec:future}

Two kinds of next step follow from this deployment: measuring the instrument further, and extending what it can be used for. The first measurement is a comparison with human examiners, which requires a new study rather than further analysis of these data (Section~\ref{sec:limitations}). A sample of recorded sittings would be graded independently by human examiners who see the submission and the transcript but not the system's score or feedback. The sample would be stratified so that the full range of scores the system produced is represented. Agreement would be reported between examiners as well as between examiner and system, since the second figure cannot be interpreted without the first. Acceptance thresholds would be fixed before any grading begins, as the criteria for this deployment were. What counts as acceptable disagreement is unsettled for human examiners too, so the design has to state its answer in advance rather than derive one afterwards.

A second measurement concerns the examination itself. Grading is already known to be stable on a fixed transcript (Section~\ref{sec:reliability}), and a student's two sittings here agreed about as closely as two oral examinations given by staff (Section~\ref{sec:disc-sittings}). Examining students twice on the same submission, with questions drawn and phrased afresh, would separate measurement error from real differences in understanding and so give the reliability of a single sitting directly. Controlled studies could also measure how far a score moves when the same question is reworded or a fixed set of questions is reordered.

A third question is how the examination shapes what students do before it. Students who know they will be asked to explain their code may read it more closely, use AI assistants differently, and come to understand their own programs better. After the exercise 3 examination, students agreed more than they had beforehand that preparing for it deepened their understanding of their code (\ref{app:reception}). That is a self-report, and a study of how students prepare would test it, for example by comparing how students work and how they perform on later tasks between cohorts or courses that do and do not examine students on their own code.

The transcripts are a further source of evidence. A qualitative reading would describe what the scores are made of, including how a student's account of their own code develops under questioning. Read across a course, the examiner's per-answer scores, grouped by the topic and dimension of each question (Section~\ref{sec:questioning}), could also show an instructor which topics and dimensions the class as a whole understood least, and they would do so without further marking.

The second kind of next step extends what the system can be used for. The same dialogue could be offered as ungraded rehearsal before the first graded sitting, after which ratings did not rise (Section~\ref{sec:disc-reception}), to measure who takes it up and whether the examination proceeds differently without stakes. It could also re-examine a corrected submission: after a failed exercise, the examiner would read the change the student made and question them on what was wrong, what changed and why it now works. Two further courses, outside the scope of this paper, have used the system since the deployment reported here: an introductory programming course, and a linear algebra course in which students scanned their handwritten work and an off-the-shelf tool converted it to text for the examiner. The first required almost no adaptation, and the second some, far less than building the system did.

\section{Conclusion}\label{sec:conclusion}

Generative AI has made working code weak evidence that a student understands it. This paper shows that a course can respond by asking every student to explain their own code, with no examiner in the room and with a grade of record at the end. To our knowledge, ATC is the first system to bring questioning anchored in the student's submission together with autonomous examination in a running course (Section~\ref{sec:related}).

The deployment did more than run. Judged against criteria fixed in advance, a 16-minute dialogue produced a score that separated students whom the course's tests could not, that was unrelated to the quality of the code they submitted yet tracked how they stood across their degree, and that, over two sittings, fell within the range of grades from semester-long courses. The examining itself, the part of an oral examination that grows with every student, was done by the system, and most of the human work that remained grew with the number of exercises rather than the number of students.

This changes what a course can ask of work done with AI. Students may use it to write their code, and their grade still rests on what they can explain about that code. The system has since been taken up by two further courses, one of them outside programming, and the comparison with human examiners described in Section~\ref{sec:future} is the next step toward using it as widely as the problem it addresses.

\appendix
\section{Implementation detail}\label{app:implementation}

This appendix describes how the system was implemented. The deployment ran on Google Cloud Platform: the application on Cloud Run, the managed container service, with examination state in a managed Postgres database and submissions and results in object storage.

\subsection*{Topic selection and structured output}

The topic selector described in Section~\ref{sec:questioning} is deterministic code, seeded separately for each sitting. Its state is kept with the rest of the sitting, as described below, so an examination can be reproduced from its record. Each agent returns its output through the provider's tool-use interface, which constrains the response to a fixed set of fields, such as a score and feedback for each answer. The system therefore never has to extract a score from free text.

\subsection*{State, access and operation}

The backend holds no examination state in memory between requests. The entire examination in progress, including the selector's state, the model message history, the transcript and the current turn, is serialized onto a single database row and reconstructed on each request. Because any instance can handle any request, the service can scale out under load, and an examination continues if the server handling it is replaced partway through. Once the examination ends, the record of the sitting holds the code examined, the full transcript, the examiner's per-answer scores and authorship ratings, the seed and the temperature, which model answered each call, and the structured outputs of the Code Reviewer and the Grader. This is the audit trail that Section~\ref{sec:goals} requires of a defensible grade.

The student's browser never reaches the backend. The student interface is a web application whose server-side layer is the only component that talks to the backend, which is private and reachable only with a short-lived credential issued by that layer. Nothing running in the browser can therefore call the agents or read stored records directly. The same application image is deployed as student, staff and quality assurance variants that differ only in runtime configuration.

Grading runs as a background task after the examination ends and can be re-run for any sitting, which allows recovery from transient failures. In the deployment reported here, each sitting that produced a grade was graded once, and no stored grade was regenerated. A staff console reports grade distributions and flagged cases, shows how often each topic was used, and exports the results table. Once the results pass a completeness check, the system emails each student the grade and the written feedback.

\section{Robustness checks}\label{app:stats}\label{app:robustness}

This appendix gives the detail behind the six checks summarized in Section~\ref{sec:robustness}, in the same order, followed by the language of the examination, three further checks and the Pearson equivalents of the main coefficients.

\paragraph{Answer length}
The oral score rises with the total number of words the student typed in the sitting, at +0.437 [+0.219, +0.608] on exercise 1 (n = 82), +0.385 [+0.230, +0.514] on exercise 2 (n = 162) and +0.373 [+0.249, +0.491] on exercise 3 (n = 215), for answers of a median of 68 words. The code quality score of the same submissions does not, at −0.016, +0.058 and +0.036 with every interval spanning zero, so length goes with the examination rather than with the quality of the student's work in general. A fuller explanation is often a better answer, but length may also be rewarded for its own sake, and these data cannot separate the two. On exercise 3, length itself correlates +0.280 [+0.154, +0.397] with grade point average (n = 213). Controlling for it, the oral score still correlates +0.349 [+0.216, +0.464] with grade point average, against +0.415 without the control on the same students, while length, controlling for the oral score, correlates +0.147 [+0.009, +0.284] with grade point average. On exercise 2 the oral score's partial coefficient is +0.343 [+0.202, +0.473] (n = 162). On exercise 1 it is +0.222 [−0.001, +0.419] (n = 80), so there the relation is not established once length is controlled. Counting only submitted answers, leaving out text cut off when time ran out, changes the partial coefficients by at most 0.013.

\paragraph{Timeouts and group}
S1 to S3 are evaluated on exercise 3, where time ran out before the last answer was submitted in 19 of the 216 sittings (Section~\ref{sec:timeouts}). Excluding those sittings gives −0.045 [−0.188, +0.095] for S1 (n = 197), +0.404 [+0.276, +0.519] for S2 and +0.444 [+0.257, +0.627] for S3 (n = 195), each on the same side of its threshold. On exercise 3, the only exercise on which both groups were assigned the AI examination, the two groups' grade distributions do not differ detectably. The median oral score is 88 in both groups (Cliff's $\delta$ = +0.02, Mann--Whitney $p$ = 0.81), the median code quality score 82 in both ($\delta$ = 0.00, $p$ = 0.99), and the median combined grade 86 in both ($\delta$ = +0.05, $p$ = 0.56). On exercises 1 and 2 each examination arm was a single group, so no such comparison exists.

\paragraph{Consistency of the code quality score}
A student's code quality scores are fairly consistent across exercises. On the same 212 pairs of sittings that give the oral figure of +0.313 in Section~\ref{sec:reliability}, the two code quality scores correlate +0.388 [+0.254, +0.503]. Exercises 1 and 3 give +0.440 [+0.193, +0.623] (n = 69), and exercises 2 and 3 give +0.443 [+0.280, +0.574] (n = 144, all students with both scores).

\paragraph{Style assessment}
The Code Reviewer's style assessment, a level from 0 to 3, relates to the two scores in opposite directions. The code quality score rises with the assessed level of machine-generated style, at +0.295, +0.376 and +0.341 on exercises 1 to 3, and the oral score falls slightly with it, at −0.076 [−0.293, +0.156], −0.248 [−0.391, −0.100] and −0.146 [−0.269, −0.019]. Holding the style level constant leaves the oral and code quality scores uncorrelated, at −0.013 [−0.238, +0.200], −0.014 [−0.169, +0.133] and −0.026 [−0.156, +0.106] on exercises 1 to 3 (n = 82, 162 and 215), against −0.035, −0.106 and −0.074 without the control on the same records, and every upper bound stays below the 0.30 that S1 sets. The style assessment is not validated as a detector (Section~\ref{sec:goals}), so these figures concern a style signal rather than the use of AI.

\paragraph{Order of sitting}
Each exercise had one topic pool for all its students, so topics recurred across sittings. Every sitting in the second half of each exercise met at least one topic already used in the first half, and the most frequent topic appeared in 19, 31 and 31 percent of the sittings on exercises 1 to 3. Because the examiner formulates each question from the student's own code, a topic does not carry its question text with it. Later sitters did not score higher in aggregate. The oral score against the order of sitting is −0.027 [−0.234, +0.194] on exercise 1 (n = 83), −0.183 [−0.328, −0.029] on exercise 2 (n = 162) and −0.097 [−0.221, +0.036] on exercise 3 (n = 216). The medians of the first and second halves are 79 and 80.5, 81 and 80, and 88 and 88.

\paragraph{Corrected code quality scores}
The recurring review error described in Section~\ref{sec:operation} affected 148 of the 216 exercise 3 code quality scores, and the course added 7 points to each of them. On the corrected scores, S1 is +0.002 [−0.128, +0.135] (n = 216) and S3 is +0.412 [+0.232, +0.580] (n = 214), so both verdicts stand. The corrected code quality score still shows no established relation to academic standing, at +0.108 [−0.030, +0.244] with grade point average (n = 214). Its consistency with the student's earlier code quality score falls from +0.388 to +0.277 [+0.133, +0.406] (n = 212).

\paragraph{Language of the examination}
Students chose the language of each examination (Section~\ref{sec:languages}). English sittings among the usable records were few (5, 14 and 20 on exercises 1 to 3), too few for a firm comparison. Their median oral scores were 80, 79 and 82, against 80, 81 and 88 in Hebrew. On exercise 3, where the gap is largest, the students who chose English also had a lower grade point average (medians 80.0 and 83.5). With grade point average controlled, the difference in oral score is not detectable (Cliff's $\delta$ = −0.197 [−0.425, +0.080], on the 19 and 195 students with a grade point average).

\paragraph{Further checks}
The agreement between a student's two oral scores, +0.313 overall (Section~\ref{sec:reliability}), is +0.287 [+0.066, +0.483] in group S (n = 69) and +0.325 [+0.164, +0.473] in group L (n = 143). Because exercise 3 oral scores run higher than those of exercises 1 and 2 (Section~\ref{sec:distributions}), the mean of two was also computed on scores standardized within each exercise, and it ranks students almost identically (Spearman +0.992 between the two versions, n = 212). The course counted a student's last graded sitting (Section~\ref{sec:booking}), whereas the analyses use the first. Using the last on exercise 3 changes the oral score of 11 students and gives −0.061 [−0.196, +0.074] for S1.

\paragraph{Pearson coefficients}
Product-moment correlations (Pearson), reported here as a robustness check (Section~\ref{sec:analysis}), are close to the rank correlations reported in the body. For the oral score against degree grade point average they are +0.411 on exercise 3 (n = 214), +0.426 on exercise 2 (n = 162), +0.329 on exercise 1 (n = 81) and +0.448 for the mean of two orals (n = 210). Pearson runs slightly below Spearman where the oral distribution is compressed, by 0.01 on exercise 3 and by 0.02 on the mean of two, and slightly above it on exercise 2, where it is not. The two statistics diverge for the code quality score and the automated test score on exercise 2, at +0.017 by rank and +0.236 by Pearson (n = 162): most students sit at 100 on the automated tests, and the Pearson figure is driven by a few low scores.

\section{Participation and retake accounting}\label{app:participation}

This appendix completes Sections~\ref{sec:participation} and~\ref{sec:retakes}. Three exercises across 247 students give 741 student-exercise cells, of which 494 were cells in which an examination was expected (Section~\ref{sec:participation}). Three sittings fall outside those 494, by students who sat an AI examination on an exercise their group was told would be autograded. One on each of exercises 1 and 2 is among the usable records. The third, by another group L student on exercise 1, whose grade on that exercise came from the automatic tests, ended before any answer was graded and was invalidated, so it carries no grade. Of the 216 students with a usable exercise 3 record, two had sat one of these examinations. Two further students, on the exercise 1 list of group S, were recorded in group L for administrative reasons and are counted in group L throughout. Neither has a usable exercise 3 record. The 21 exercise 3 students who submitted but had not sat by the export cut held no grade for the exercise. Bookings and missed bookings were not recorded in the data exported for analysis, so these students cannot be split into those who never booked and those who booked and did not attend.

Staff invalidated a sitting to permit a further attempt (Section~\ref{sec:booking}), so an invalidated sitting can still carry a grade, and 56 of the 68 invalidated sittings did. A kept sitting, as in Section~\ref{sec:retakes}, is any sitting except one that carried no final grade or ended in the confirmed technical fault, so those 56 are kept. The analyses that test the success criteria use the first kept sitting in each cell. A sitting whose earlier sitting carried no grade therefore counts as the first, and the 10 cells in that position do not count as retakes. On exercise 3, 14 of the 216 usable records are graded sittings that were later invalidated at the student's request. In 9 of those cells the student sat again, and in 5 no further sitting took place. Together with 2 cells in which a timed-out sitting was followed by another, this gives the 11 exercise 3 cells with more than one kept sitting.

Of the 461 student-exercise cells holding at least one kept sitting, 403 hold exactly one, 46 hold two and 12 hold three. The 58 cells holding more than one divide across the exercises as 18 on exercise 1, 29 on exercise 2 and 11 on exercise 3. Of the 54 students appearing in them, 50 retook on a single exercise and 4 on two. From the first to the last kept sitting in these cells, the change in oral score (Section~\ref{sec:retakes}) ranges from −20.0 to +38.0 points.

The highest difficulty a sitting reached is compared over the same 58 cells (Table~\ref{tab:difficulty}). Because every examination opens at medium, that level is either medium or hard, although 111 questions in the usable sittings were asked at the easy level. It rose from medium to hard in 25 cells and fell from hard to medium in 3. As with the oral score (Section~\ref{sec:retakes}), the rise does not show that students improved: a retake usually followed a sitting that went badly, and part of a bad result is chance.

\begin{table}[t]
\centering
\small
\caption{Highest difficulty reached on the first and last kept sitting, for the 58 cells with more than one kept sitting.}
\label{tab:difficulty}
\begin{adjustbox}{max width=\textwidth}
\begin{tabular}{lcc}
\toprule
First sitting & Last sitting & Cells \\
\midrule
medium & hard & 25 \\
hard & medium & 3 \\
hard & hard & 15 \\
medium & medium & 15 \\
 & Total & 58 \\
\bottomrule
\end{tabular}
\end{adjustbox}
\end{table}

\section{Survey instrument}\label{app:survey}

The three forms were sent as described in Section~\ref{sec:surveys}: Form 1 or Form 2, according to the student's arm, before each exercise was graded, and Form 3 after each AI examination. Eight students answered both Form 1 and Form 2 on the same exercise, seven of them on exercise 1. The English below is a literal translation. Every scale item used the same instruction and scale: ``Rate how far you agree with each statement'', from 1 = do not agree at all to 7 = completely agree. Items T0 to L3 on Forms 1 and 3 form the eight-item scale analyzed in \ref{app:reception}, and T0 and T01, worded identically on Forms 1 and 2 apart from the name of the mode, are the items behind H1.

\subsection*{Form 1: told-AI arm, before the examination (10 items)}

\begin{center}
\small
\begin{tabular}{@{}l p{0.86\textwidth}@{}}
\toprule
\multicolumn{2}{@{}l}{\textit{Part A: Expectations of the assessment}} \\
\midrule
T0 & I believe that the grade I will receive from an AI-based assessment system will faithfully reflect my level of understanding of the material. \\
T01 & I believe that the grade I will receive from an AI-based assessment system will faithfully reflect my mastery of the material. \\
T1 & I believe that having the exercise assessed by an AI-based assessment system will allow me to present my knowledge in full. \\
T2 & I believe that an AI-based assessment system will ask the right questions about my code. \\
T3 & I believe in the accuracy of the feedback that an AI-based assessment system will give me. \\
\midrule
\multicolumn{2}{@{}l}{\textit{Part B: Expected contribution to learning}} \\
\midrule
L1 & I believe that the feedback from an AI-based assessment system will contribute to my learning process in the course. \\
L2 & I believe that preparing for assessment by an AI-based assessment system will lead me to a deeper understanding of my code. \\
L3 & I believe that the dialogue with an AI-based assessment system will deepen my understanding of the code. \\
\midrule
\multicolumn{2}{@{}l}{\textit{Part C: Assessor preference}} \\
\midrule
P1 & In your view, who is best suited to assess programming exercises in the course? \\
P2 & By which method would you prefer your assignment to be assessed in the course? \\
\bottomrule
\end{tabular}
\end{center}

\subsection*{Form 2: told-autograde arm, before grading (8 items)}

\begin{center}
\small
\begin{tabular}{@{}l p{0.86\textwidth}@{}}
\toprule
\multicolumn{2}{@{}l}{\textit{Part A: Expectations of the assessment}} \\
\midrule
T0 & I believe that the grade I will receive from the automatic assessment system will faithfully reflect my level of understanding of the material. \\
T01 & I believe that the grade I will receive from the automatic assessment system will faithfully reflect my mastery of the material. \\
T1 & I believe that having the exercise assessed by the automatic assessment system will reflect the correctness of the solution I submitted. \\
T2 & I believe that the automatic assessment system will test my code on a wide range of cases. \\
T3 & I believe in the accuracy of the grade that the automatic assessment system will give me. \\
\midrule
\multicolumn{2}{@{}l}{\textit{Part B: Expected contribution to learning}} \\
\midrule
L1 & I believe that the results from the automatic assessment system will contribute to my learning process in the course. \\
\midrule
\multicolumn{2}{@{}l}{\textit{Part C: Assessor preference}} \\
\midrule
P1 & In your view, who is best suited to assess programming exercises in the course? \\
P2 & By which method would you prefer your assignment to be assessed in the course? \\
\bottomrule
\end{tabular}
\end{center}

\subsection*{Form 3: told-AI arm, after the examination (10 items)}

\begin{center}
\small
\begin{tabular}{@{}l p{0.86\textwidth}@{}}
\toprule
\multicolumn{2}{@{}l}{\textit{Part A: The assessment experience}} \\
\midrule
T0 & The grade I received from an AI-based assessment system faithfully reflected my level of understanding of the material. \\
T01 & The grade I received from an AI-based assessment system faithfully reflected my mastery of the material. \\
T1 & Having the exercise assessed by an AI-based assessment system allowed me to present my knowledge in full. \\
T2 & An AI-based assessment system asked the right questions about my code. \\
T3 & I believe in the accuracy of the feedback that an AI-based assessment system gave me. \\
\midrule
\multicolumn{2}{@{}l}{\textit{Part B: Contribution to learning}} \\
\midrule
L1 & The feedback from an AI-based assessment system contributed to my learning process in the course. \\
L2 & Preparing for assessment by an AI-based assessment system led me to a deeper understanding of my code. \\
L3 & The dialogue with an AI-based assessment system deepened my understanding of the code. \\
\midrule
\multicolumn{2}{@{}l}{\textit{Part C: Assessor preference (after the experience)}} \\
\midrule
P1 & After the experience you have had, who do you think is best suited to assess programming exercises? \\
P2 & After the experience you have had, by which method would you prefer your assignment to be assessed in the future? \\
\bottomrule
\end{tabular}
\end{center}

\subsection*{Answer options for the two preference items (identical on all three forms)}

P1: the automatic assessment system only; an AI-based assessment system only; a combination of the two methods.

P2: assessment by the automatic assessment system; assessment by an AI-based assessment system; a combination of the two methods; I have no preference.

\section{Student reception: decision rules and full results}\label{app:reception}

This appendix completes Section~\ref{sec:reception}, and Table~\ref{tab:hypotheses} summarizes the ten hypotheses.

\begin{table}[t]
\centering
\small
\caption{The ten survey hypotheses. Item codes are those of the forms in \ref{app:survey}. Primary hypotheses carry 98.33\% intervals (Bonferroni for three tests), exploratory ones 95\% intervals. $\delta$ is Cliff's delta between two sets of respondents, $r$ the matched-pairs rank-biserial correlation, and the net switch the share of students who moved to an option that includes the AI examination minus the share who moved away from one.}
\label{tab:hypotheses}
\begin{adjustbox}{max width=\textwidth}
\begin{tabular}{>{\raggedright\arraybackslash}p{3.6cm}l>{\raggedright\arraybackslash}p{4.2cm}lll}
\toprule
Hypothesis & Status & Contrast & Estimate [interval] & n & Verdict \\
\midrule
H1 naive expectation gap & primary & $\delta$ on T0 and T01, autograder against AI pre form, exercise 1 & $+0.305$ [$+0.093$, $+0.490$] & 103 and 74 & supported \\
H2 question-appropriateness surprise & exploratory & $r$ on T2, post against pre, group L at exercise 2 & $+0.355$ [$+0.126$, $+0.544$] & 122 pairs & supported \\
H3 exercise-three experience gain & primary & $r$ on T0--L3, post against pre, exercise 3 & $+0.489$ [$+0.214$, $+0.699$] & 107 pairs & supported \\
H4 retention-support shift & primary & net switch on P2, exercise 3 & $+0.159$ [$+0.034$, $+0.273$] & 88 pairs & supported \\
H5 and H10 order pair & exploratory & $\delta$ on T0--L3 after the examination, group S against L, exercise 3, margin $\pm 0.15$ & $-0.019$ [$-0.264$, $+0.223$] & 35 and 77 & neither \\
H6 learning-contribution expectation gap & exploratory & $\delta$ on L1, autograder against AI pre form, exercise 1 & $+0.052$ [$-0.132$, $+0.215$] & 103 and 74 & not supported \\
H7 first-encounter gain & exploratory & $r$ on T0--L3, post against pre, group S at exercise 1 & $-0.349$ [$-0.603$, $-0.064$] & 63 pairs & not supported (reversed) \\
H8 broad post-autograder gain & exploratory & $r$ on T0--L3, post against pre, group L at exercise 2 & $+0.042$ [$-0.179$, $+0.250$] & 122 pairs & not supported \\
H9 accuracy-perception gain & exploratory & $r$ on T3, post against pre, exercise 3 & $+0.195$ [$-0.077$, $+0.429$] & 107 pairs & not supported \\
\bottomrule
\end{tabular}
\end{adjustbox}
\end{table}

\paragraph{Forms and records}
The three forms are given in \ref{app:survey}. Because there is no post-examination form for the autograder, every comparison between the two modes sets expectations against expectations, while every comparison before and after the AI examination is within student, on items whose wording matches one to one. Six scale items appear on both pre-examination forms. Two are worded identically apart from the name of the mode (T0 and T01), two were adapted to each mode because an autograder returns no feedback (T3 and L1), and two ask about a different aspect of each mode (T1 and T2). Some students submitted the same form on the same exercise more than once: 78 students did so, in 90 instances. The earliest record is used, and the latest is used in a pre-specified robustness check. The four hypotheses that compare the two arms or the two groups (H1, H5, H6 and H10) exclude the five students whose experience on exercise 1 or 2 departed from their group's arm: the three who sat under the other group's arm and the two recorded in group L for administrative reasons (\ref{app:participation}), and including them changes no verdict.

Response was high on the pre-examination AI form (89, 94 and 83 percent of eligible students on exercises 1 to 3) and on the post-examination form at exercises 1 and 2 (81 and 79 percent of those who sat). It was lower on the pre-examination autograder form (60 and 32 percent) and on the post-examination form at exercise 3 (52 percent). Complete pre-post pairs cover 77, 75 and 50 percent of the students who sat, so the exercise-three experience gain rests on about half of that exercise's cohort. The students who answered both forms also differ from those who did not: on exercise 3 their oral scores were higher (medians 91 and 87, Cliff's delta +0.167 [+0.012, +0.317]).

\paragraph{Decision rules}
The hypotheses, their measures and their thresholds were written after the deployment and before any response was inspected (Section~\ref{sec:analysis}). A hypothesis of a difference is supported when its interval, at the level stated for it, excludes zero and its point estimate runs in the predicted direction. H5, a hypothesis of equivalence, is supported when its interval lies within ±0.15 on Cliff's delta. Two further conditions apply: at least 30 complete pairs for a contrast within students and 25 respondents per group for a contrast between groups, and an internal consistency of at least 0.70 for any composite.

\paragraph{Primary hypotheses}
The naive expectation gap (H1) asked whether students, before experiencing either mode, would expect the autograder's grade to reflect their understanding and mastery better than the AI examination's. It is tested at exercise 1, the only point at which no student had experienced either mode, on the two identically worded items (T0 and T01). The two-item composite gives a Cliff's delta of +0.305, with a 98.33\% interval of [+0.093, +0.490] and p = 0.00046, on 103 respondents to the autograder form and 74 to the AI form. Each item excludes zero on its own: whether the grade reflects understanding, at +0.299 [+0.092, +0.483], and whether it reflects mastery, at +0.273 [+0.066, +0.460]. The adapted accuracy item, which does not enter the decision, gives +0.403 [+0.201, +0.575], but it asks about the accuracy of the AI examiner's feedback and of the autograder's grade, so a difference on it may reflect those two objects rather than the two modes. Seven students answered both pre-examination forms at exercise 1 and so appear on both sides of the contrast. Excluding them gives +0.344 [+0.125, +0.531] on 96 and 67 respondents, and a bootstrap that keeps each such student's two records together gives [+0.089, +0.491] around the same +0.305.

The exercise-three experience gain (H3) asked whether students would rate the AI examination more positively after the exercise 3 examination than before it. By then almost every student had experienced both modes and was meeting the AI examination for the second time. Across the eight-item composite (T0 to L3), the matched-pairs rank-biserial correlation is +0.489, with a 98.33\% interval of [+0.214, +0.699] and $p = 3.1 \times 10^{-5}$, on 107 complete pairs, and all eight item estimates are positive. The retention-support shift (H4) asked whether that examination would increase support for keeping an AI-based component in grading, measured by the preference item P2. The net switch, the share of students who moved to an option that includes the AI examination minus the share who moved away from one, is +0.159, with a 98.33\% interval of [+0.034, +0.273] and p = 0.0026, on the 88 of 107 pairs left after setting aside respondents who chose no preference at either end. The pre-specified alternative coding, which counts no preference as not including AI, agrees, at +0.140 [+0.019, +0.252] on 107 pairs. All three primary hypotheses meet the further conditions: each contrast exceeds its minimum sample, and the composites reach an internal consistency of 0.968 for the two-item composite (Spearman--Brown on the polychoric correlation) and 0.960 and 0.955 for the two eight-item composites (ordinal alpha).

On the exercise 3 composite, five of the eight items have 98.33\% intervals that exclude zero: whether the examiner asks the right questions, whether the grade reflects mastery, whether it reflects understanding, whether the student could present their knowledge fully, and whether preparing for the examination deepened their understanding. The three that do not are whether the feedback helps learning, whether the dialogue deepens understanding, and the accuracy of the feedback. Because the eight items correlate highly with one another (mean inter-item Spearman near 0.69), they do not measure eight distinct attitudes, so the number of items that moved reflects the size of the shift rather than its breadth.

Three features qualify these results. H1 compares the respondents to the two forms, who come almost entirely from different lecture groups, and the groups differ in size and in degree-track composition, so the contrast between the two modes is also a contrast between the groups. H3 and H4 were answered shortly after a grade from the system, at a second encounter with the format, and part of the movement may reflect satisfaction with that grade. They also rest on the students who answered both forms, whose oral scores were higher than those of the students who did not.

\paragraph{Exploratory hypotheses}
Of the seven exploratory hypotheses, one is supported. Students meeting the AI examination after the autograder, at exercise 2, found its questions (T2) more appropriate than they had expected (H2, the question-appropriateness surprise), at +0.355 [+0.126, +0.544] on 122 pairs. One result runs against its prediction. Students meeting the AI examination first, at exercise 1, rated it lower afterwards than they had expected on all eight items, four of them with intervals below zero (H7, the first-encounter gain), at −0.349 [−0.603, −0.064] on 63 pairs, with 36 students moving down and 23 up. Satisfaction with a grade would push ratings after the examination up rather than down, so it does not explain this result, but the novelty of the format and the wider spread of exercise 1 scores may. The same eight items were measured in three cells, giving +0.489 at exercise 3, −0.349 for group S at exercise 1 and +0.042 for group L at exercise 2, so the exercise 3 gain does not repeat at either group's first encounter with the AI examination.

Three further exploratory hypotheses are not supported. Before either mode, students showed no detectable difference in how much they expected each to contribute to their learning (L1; H6, the learning-contribution expectation gap), at +0.052 [−0.132, +0.215] on the same 103 and 74 respondents, although this rests on a single item worded differently for each mode. Students meeting the AI examination after the autograder did not rate it more favorably overall (H8, the broad post-autograder gain), at +0.042 [−0.179, +0.250] on 122 pairs, although the appropriateness of its questions (H2) is one of the eight items in that composite. Perceived accuracy (T3) did not change detectably at exercise 3 (H9, the accuracy-perception gain), at +0.195 [−0.077, +0.429] on 107 pairs. The accuracy item is also one of the three items in the H3 composite whose interval includes zero, so H9 adds no independent evidence.

The order pair tested whether the order in which the two groups had met the two modes left a mark on their ratings after the exercise 3 examination, across T0 to L3. H5 predicted that the two groups' ratings would be equivalent, within a margin of ±0.15 on Cliff's delta, and H10 that they would differ. The difference, −0.019 [−0.264, +0.223] on 35 and 77 respondents, is too wide to support either hypothesis.

\paragraph{Robustness}
Using each student's latest record instead of the earliest, the pre-specified robustness check, leaves every verdict unchanged. It changes the estimates by at most 0.117, and by about 0.04 at the median. The duplicated records concentrate on the pre-examination AI form and on exercise 2, so the choice of record matters more for some comparisons than for others, but it moves no verdict.

\section*{Funding}

This research did not receive any specific grant from funding agencies in the public, commercial, or not-for-profit sectors.

\section*{Declaration of competing interest}

The authors declare that they have no known competing financial interests or personal relationships that could have appeared to influence the work reported in this paper.

\section*{Data availability}

The anonymized per-student data analyzed in this study are available as ancillary files with this arXiv submission: the oral, code quality and automated test scores, degree grade point average, and the responses to the survey's scale and preference items. The instructions given to the examiner, Code Reviewer and Grader agents are not released, at the request of the authors' institution. Section~\ref{sec:design} and \ref{app:implementation} describe what each agent receives and what it produces.

\section*{Declaration of generative AI and AI-assisted technologies in the manuscript preparation process}

During the preparation of this work the authors used Claude (Anthropic) in order to improve phrasing and language clarity. After using this tool, the authors reviewed and edited the content as needed and take full responsibility for the content of the published article.

\bibliographystyle{elsarticle-harv}
\bibliography{references}

\end{document}